\documentclass{emulateapj}

\newcommand {\lya}    {Ly$\alpha$}   
\newcommand {\lyb}    {Ly$\beta$}    
\newcommand {\lyg}    {Ly$\gamma$}
\newcommand {\lyd}    {Ly$\delta$}
\newcommand {\lye}    {Ly$\epsilon$}
\newcommand {\lyz}    {Ly$\zeta$}
\newcommand {\lyeta}  {Ly$\eta$}
\newcommand {\HH}     {H$_2$}        
\newcommand {\HI}     {\ion{H}{1}}   
\newcommand {\OVI}    {\ion{O}{6}}   
\newcommand {\CIII}   {\ion{C}{3}}   

\newcommand {\kms}    {km~s$^{-1}$}
\newcommand {\NOVI}   {$N_{\rm OVI}$}
\newcommand {\NHI}    {$N_{\rm HI}$}
\newcommand {\NCIII}  {$N_{\rm CIII}$}
\newcommand {\tnmc}{\tablenotemark{c}}
\newcommand {\lam}    {$\lambda$}
\newcommand {\FUSE}  {{\it FUSE}} 

\begin{document}

\title{The Low-\lowercase{$z$} Intergalactic Medium. II. \lyb, O\,VI, and C\,III Forest}

\author{Charles W. Danforth, J. Michael Shull, Jessica L. Rosenberg\altaffilmark{1}, \& John T. Stocke}
\affil{Center for Astrophysics and Space Astronomy, Dept. of Astrophysial \& 
Planetary Sciences, University of Colorado, Boulder, CO 80309; danforth@casa.colorado.edu, mshull@casa.colorado.edu, stocke@casa.colorado.edu}
\altaffiltext{1}{NSF Astronomy and Astrophysics Postdoctoral Fellow, currently at Harvard-Smithsonian Center for Astrophysics; \\ jlrosenberg@cfa.harvard.edu}


\begin{abstract}
We present the results of a large survey of \HI, \OVI, and \CIII\ absorption lines in the low-redshift ($z<0.3$) intergalactic medium (IGM).  We begin with 171 strong \lya\ absorption lines ($W_{\lambda} \geq 80$~m\AA) in 31 AGN sight lines studied with the {\it Hubble Space Telescope} and measure corresponding absorption from higher-order Lyman lines with \FUSE.  Higher-order Lyman lines are used to determine \NHI\ and $b_{\rm HI}$ accurately through a curve-of-growth (COG) analysis.  We find that the number of \HI\ absorbers per column density bin is a power-law distribution, $d{\cal N}/dN_{\rm HI}\propto N_{\rm HI}^{-\beta}$, with $\beta_{\rm HI}=1.68\pm0.11$.  We made 40 detections of \OVI\ \lam\lam1032,1038 and 30 detections of \CIII\ \lam977 out of 129 and 148 potential absorbers, respectively.  The column density distribution of \CIII\ absorbers has $\beta_{\rm CIII} = 1.68\pm0.04$, similar to $\beta_{\rm HI}$ but not as steep as $\beta_{\rm OVI} = 2.1 \pm 0.1$.  From the absorption-line frequency, $d{\cal N}_{\rm CIII}/dz=12^{+3}_{-2}$ for $W_{\lambda}({\rm CIII})>30$~m\AA, we calculate a typical IGM absorber size $r_0\sim400$~kpc, similar to scales derived by other means.  The COG-derived $b$-values show that \HI\ samples material with $T<10^5$~K, incompatible with a hot IGM phase.  By calculating a grid of CLOUDY models of IGM absorbers with a range of collisional and photoionization parameters, we find it difficult to simultaneously account for the \OVI\ and \CIII\ observations with a single phase.  Instead, the observations require a multiphase IGM in which \HI\ and \CIII\ arise in photoionized regions, while \OVI\ is produced primarily through shocks.  From the multiphase ratio \NHI/\NCIII, we infer the IGM metallicity to be $Z_C = 0.12\,Z_\sun$, similar to our previous estimate of $Z_O = 0.09\,Z_\sun$ from \OVI.  
\end{abstract}



\section{Introduction} The study of the intergalactic medium (IGM) is crucial for understanding the structure and evolution of galaxies.  Much has been learned from the distribution of visible galaxies in large-scale structure, but a large fraction of the baryonic mass of the universe still resides in the IGM \citep{CenOstriker99a,CenOstriker01,Dave99,Dave01}, even at low redshift \citep{Stocke04}.  These intergalactic absorbers are likely composed of primordial material left over from the big bang and processed material ejected from galaxies.  The identification of these absorbers and their origins will help to constrain both the evolution of primordial gas in the universe as well as galactic outflows and metal processing.  By comparing local, low-redshift IGM absorbers with those in the high-redshift universe, we can watch the evolution of these processes through cosmic time.

The IGM is best probed with absorption-line studies using distant continuum sources such as quasars and AGN.  \HI\ is best detected through Lyman line absorption in the rest-frame far ultraviolet (FUV; 912--1216~\AA), particularly \lya\ at 1216~\AA.  Ironically, much analysis has been done of the \lya\ forest and metal lines in the distant universe ($2<z<5$), but comparatively little exists for the local universe ($z<1$).  To fill this gap in our knowledge, we must examine the IGM in the FUV.

The \lya\ line is arguably the most important diagnostic of the IGM, but it has inherent limitations.  Strong lines exhibit saturation and blending from weaker, unresolved components.  Profile fits to the \lya\ line tend to underestimate \HI\ column density and overestimate the Doppler width \citep{Shull00}.  For more accurate analysis, weaker Lyman lines such as \lyb\ are required in conjunction with \lya.  The trade-off is that weaker lines are more difficult to detect and, at low redshift, lie in a wavelength range complicated with \HH\ and ionic lines from the local interstellar medium (ISM).

The {\it Hubble Space Telescope} (HST) has housed a pair of ultraviolet instruments, the {\it Goddard High Resolution Spectrograph} (GHRS) and the {\it Space Telescope Imaging Spectrograph} (STIS), which have brought UV spectroscopy into a golden age with high-resolution spectrographic capabilities down to $\sim$1150 \AA.  The {\it Far Ultraviolet Spectroscopic Explorer} (\FUSE) complements the capabilities of HST, providing coverage of the 905--1187 \AA\ band with the higher Lyman lines missed by HST at $z<0.12$ as well as the important diagnostic lines of \OVI\ $\lambda\lambda1032,1038$ and \CIII\ $\lambda977$ out to modest redshifts \citep{Moos00}.

These spectrographs have compiled a sizeable catalog of IGM absorbers along extragalactic sight lines.  Absorption systems in the IGM can be identified using strong \lya\ lines in the relatively uncomplicated spectral region of GHRS and STIS data.  Once located, the higher Lyman lines and metal diagnostics covered by \FUSE\ give a much more complete picture of the system.

In this study, we use published catalogs of \lya\ absorbers in extragalactic sight lines and search for counterpart absorption in the higher Lyman lines (\lyb, \lyg, \lyd, etc) to more accurately determine the column density \NHI\ and Doppler line width $b_{\rm HI}$ for neutral hydrogen in the IGM.  Furthermore, we look for counterpart absorption in probes of the crucial warm-hot ionized medium (WHIM), \OVI\ $\lambda\lambda1031.926,1037.617$ and possibly \CIII\ $\lambda977.020$.  \OVI\ is present in gas at $10^{5-6}$ K and is thought to arise from shock ionization perhaps with some contribution from hard, ionizing external radiation in low density gas.  The \HI\ absorption can be seen in the \FUSE\ band at $0.003 <z <0.3$, while \OVI\ and \CIII\ can be measured between $0<z<0.15$ and $0 < z <0.21$, respectively.  We use these lines to gain better understanding of how gas becomes ionized in the low-redshift universe.

We gave an overview of the project and some of the more important cosmological results in Danforth \& Shull (2005; henceforth Paper~I).  In this paper, we describe in complete detail our catalog of 171 \lya\ absorbers in sight lines toward 31 AGN, including several dozen new \lya\ absorbers.  We also describe our survey of \lyb\ and \CIII\ absorbers.  We present our criteria for selection of sight lines and absorber in \S~2 along with a list of previously unpublished \lya\ absorbers.  Our analysis methods and results are presented in \S~3.  In \S~4 we discuss the importance of multiple Lyman lines to accurate \HI\ measurements, analyze the distribution of \CIII\ detections, discuss new evidence for a multiphase IGM, and investigate the metallicity of the IGM.  We summarize our conclusions in \S~5.

\begin{deluxetable*}{lcllcllccr}
\tabletypesize{\footnotesize}
\tablecolumns{10} 
\tablewidth{0pt} 
\tablecaption{{\it FUSE} IGM Sight Lines}
\tablehead{\colhead{Target}                                    &
	   \colhead{Alternate}                                 &
           \colhead{RA\,(J2000)}                               &
	   \colhead{Dec\,(J2000)}                              & 
           \colhead{AGN}                                       &
	   \colhead{$z\rm_{AGN}$}                              &
           \colhead{$\Delta z_{\rm Ly\alpha}$\tablenotemark{a}}&
           \colhead{$\cal N\rm_{abs}$\tablenotemark{b}}        &
           \colhead{FUSE}                                      &
	   \colhead{Exp.}                                      \\
	   \colhead{AGN}                                       &        
	   \colhead{Name}                                      &  
	   \colhead{{h}\phn{m}\phn{s}}                         &
	   \colhead{\phn{\arcdeg}~\phn{\arcmin}~\phn{\arcsec}} &
	   \colhead{type}                                      & 	       
	   \colhead{}                                          & 	    
           \colhead{}                                          &
	   \colhead{}                & 
	   \colhead{ID}                                        & 
	   \colhead{(ksec)}                                    } 
\startdata
Mrk\,335     & PG\,0003$+$199 & 00 06 19.5 & $+$20 12 10.3 & Syf1  & 0.0256 & 0.0159 &   3  &P10102        &  97.0 \\
I\,Zw1       & PG\,0050$+$124 & 00 53 34.9 & $+$12 41 36.0 & Gal   & 0.0607 & 0.0298 &   3  &P11101        &  38.6 \\
Ton\,S180    & HE\,0054$-$2239& 00 57 20.0 & $-$22 22 59.3 & Syf1.2& 0.0620 & 0.0570 &   6  &P10105        &  16.6 \\
Fairall\,9   & \nodata        & 01 23 46.0 & $-$58 48 23.8 & Syf1  & 0.0461 & 0.0378 &   1  &P10106        &  38.9 \\
HE\,0226$-$4110& \nodata      & 02 28 15.2 & $-$40 57 16   & QSO   & 0.495  & 0.4061 &  11  &multiple\tnmc &  33.2 \\
NGC\,985     & Mrk\,1048      & 02 34 37.8 & $-$08 47 15.6 & Syf1  & 0.0431 & 0.0381 &   0  &P10109        &  68.0 \\
PKS\,0405$-$12 & \nodata      & 04 07 48.2 & $-$12 11 31.5 & QSO   & 0.574  & 0.4061 &   7  &B08701        &  71.1 \\
Akn\,120     & Mrk\,1095      & 05 16 11.4 & $-$00 08 59.4 & Syf1  & 0.0331 & 0.0225 &   1  &P10112        &  56.2 \\
VII\,Zw118   & \nodata        & 07 07 13.1 & $+$64 35 58.8 & Syf1  & 0.0797 & 0.0266 &   1  &multiple\tnmc & 198.6 \\
PG\,0804$+$761 & \nodata      & 08 10 58.5 & $+$76 02 41.9 & QSO   & 0.1000 & 0.0686 &   2  &multiple\tnmc & 174.0 \\
Ton\,951     & PG\,0844$+$349 & 08 47 42.5 & $+$34 45 03.5 & QSO   & 0.064  & 0.0590 &   0  &P10120        &  31.9 \\
PG\,0953$+$414 & \nodata      & 09 56 52.8 & $+$41 15 25.7 & QSO?  & 0.239  & 0.2340 &  15  &P10122        &  72.1 \\
Mrk\,421     &  \nodata       & 11 04 27.3 & $+$38 12 32.0 & Blazar& 0.0300 & 0.0203 &   1  &multiple\tnmc &  83.9 \\
PG\,1116$+$215 & Ton\,1388    & 11 19 08.7 & $+$21 19 18.2 & QSO   & 0.1763 & 0.0686 &  11  &P10131        &  77.0 \\
PG\,1211$+$143 & \nodata      & 12 14 17.6 & $+$14 03 12.7 & Syf   & 0.0809 & 0.0686 &  12  &P10720        &  52.3 \\
3C\,273      & PG\,1226$+$023 & 12 29 06.7 & $+$02 03 08.9 & QSO   & 0.1583 & 0.1533 &   8  &P10135        &  42.3 \\
PG\,1259$+$593 & \nodata      & 13 01 13.1 & $+$59 02 05.7 & QSO   & 0.472  & 0.4061 &  20  &P10801        & 668.3 \\
PKS\,1302$-$102&PG\,1302$-$102& 13 05 32.8 & $-$10 33 22.0 & QSO   & 0.286  & 0.2810 &  18  &P10802        & 142.7 \\
Mrk\,279     & PG\,1351$+$695 & 13 53 03.4 & $+$69 18 29.9 & Syf1  & 0.0306 & 0.0200 &   0  &multiple\tnmc & 228.5 \\
Mrk\,1383    & PG\,1426$+$015 & 14 29 06.6 & $+$01 17 06.6 & Syf1  & 0.0865 & 0.0686 &   2  &multiple\tnmc &  63.5 \\
Mrk\,817     & PG\,1434$+$590 & 14 36 22.1 & $+$58 47 39.5 & Syf1.5& 0.0313 & 0.0206 &   1  &P10804        & 189.9 \\
Mrk\,478     & PG\,1440$+$356 & 14 42 07.5 & $+$35 26 22.9 & Gal   & 0.0791 & 0.0686 &   3  &P11109        &  14.2 \\
Mrk\,290     & PG\,1534$+$580 & 15 35 52.4 & $+$57 54 09.3 & Syf1  & 0.0296 & 0.0114 &   0  &P10729        &  12.8 \\
Mrk\,876     & PG\,1613$+$658 & 16 13 57.2 & $+$65 43 10   & QSO   & 0.129  & 0.0266 &   2  &P10731        &  46.0 \\
H\,1821$+$643  & \nodata      & 18 21 57.3 & $+$64 20 36.3 & Syf1  & 0.2968 & 0.2810 &  16  &P10164        & 132.3 \\
PKS\,2005$-$489& \nodata      & 20 09 25.4 & $-$48 49 53.9 & BLLac & 0.0710 & 0.0660 &   4  &multiple\tnmc &  49.2 \\
Mrk\,509     & \nodata        & 20 44 09.7 & $-$10 43 24.7 & Syf1  & 0.0344 & 0.0263 &   1  &P10806        &  62.3 \\
II\,Zw136    & PG\,2130$+$099 & 21 32 27.8 & $+$10 08 19.4 & Syf1  & 0.0630 & 0.0580 &   1  &P10183        &  22.7 \\
PHL\,1811    & \nodata        & 21 55 01.6 & $-$09 22 26.0 & Syf1  & 0.192  & 0.1870 &  14  &multiple\tnmc &  75.0 \\
PKS\,2155$-$304& \nodata      & 21 58 52.1 & $-$30 13 32.3 & BLLac & 0.1165 & 0.0585 &   8  &P10807        & 123.2 \\
MR\,2251$-$178 & \nodata      & 22 54 05.8 & $-$17 34 55.0 & Gal   & 0.0644 & 0.0594 &   2  &P11110        &  54.1 
\enddata      
\tablenotetext{a}{Redshift pathlength $\Delta z$ surveyed for \lya\ absorption.  We only use absorbers at $z<0.3$ in this work.}
\tablenotetext{b}{$\cal N_{\rm abs}$ is the number of \lya\ absorbers $W_\lambda>80$~m\AA\ at $z_{\rm abs}<0.3$.}
\tablenotetext{c}{Multiple FUSE observations are as follows: 
		HE\,0226$-$4110=P10191, P20713; 
		VII\,Zw118=P10116, S60113; 
		PG\,0804$+$761=P10119, S60110; 
		Mrk\,421=P10129, Z01001; 
		Mrk\,279=P10803, C09002, D15401; 
		Mrk\,1383=P10148, P26701; 
		PKS\,2005$-$489=P10738, C14903; 
		PHL\,1811=P10810, P20711}
\end{deluxetable*}

\section{Sight Lines and Absorbers}

The catalog of \FUSE\ observations of extragalactic sight lines is large enough that a statistical study of \OVI\ in the low-redshift IGM is feasible.  We searched the catalog of observations for sight lines with previous quasar absorption-line studies.  The majority of our target list was obtained from the Colorado surveys based on GHRS and STIS grating observations of 30 AGN \citep{Penton1,Penton2,Penton3,Penton4}.  Six of these targets lack \FUSE\ data of sufficient quality to be included in our survey.  The target list was augmented by additional studies of these sight lines from the literature, using all available moderate and high-resolution UV data.  We analyzed archival STIS/E140M echelle data for six additional sight lines as described below.  Our final sample consists of 31 sight lines with reasonable quality \FUSE\ data and a known set of \lya\ absorbers (or a known lack of \lya\ absorbers) over at least part of their path length (Table~1).  The total pathlength surveyed for \lya\ absorbers is noted in column 7 of Table~1.

One of our goals was to rigorously determine the \HI\ column density for the IGM absorbers via curves of growth.  Different authors use different techniques to determine column densities and $b$-values, resulting in a heterogenous sample of absorber information.  For this reason, the primary information we take from the \lya\ surveys is absorber redshift $z_{\rm abs}$ and \lya\ equivalent width $W_{\rm Ly\alpha}$, both of which are measured directly from the data and are not subject to analytical interpretation.

In six cases, we used archival STIS/E140M data and performed our own \lya\ absorber search.  The data were smoothed by three pixels to better match the instrumental resolution.  For four targets (PKS\,0405$-$123, PG\,0953$+$415, PG\,1259$+$593, and PKS\,1302$-$102), our analysis was carried out as follows: (1) the major absorption and emission line regions were marked by hand for exclusion; (2) the excluded regions of the spectrum were then replaced with a linear interpolation over ten pixels on either side of the excluded region; (3) a rough smoothing over 100 pixels was applied to the resulting continuum;, finally (4) a cubic-spline linear interpolation was derived from all of the continuum data points that were less than 2$\sigma$ from the smoothed continuum value.  This linear interpolation over points deviating by less than 2$\sigma$ from the smoothed value for the continuum was the fit used in the processing of the data.  The one exception was for lines that fell on the damping wings of Galactic \lya; in these cases a local fit to the continuum was performed.

The measurement of the absorption lines in the E140M data was performed using the program VPFIT\footnote{VPFIT was written by R.~F. Carswell, J.~K. Webb, M.~J. Irwin, \& A.~J. Cooke.  More information is available at \tt http://www.ast.cam.ac.uk/$\sim$rfc/vpfit.html} which performs Voigt profile fitting for each of the lines.  The line fitting was performed interactively for each of the lines.  In general, we were conservative about adding multiple components to the fit since the fit can almost always be improved by adding these components even if it is not justified by an F-test.  We did, however, add multiple components to the \lya\ fits in cases where the higher Lyman lines or metal lines for the same absorber indicated a multi-component structure.  In some of these cases, we use the information derived from the high-order lines to fix the positions and/or the $b$-values for the \lya\ line fits.

In two other cases, HE\,0226$-$4110 and PHL\,1811, we analyzed the STIS/E140M data using an IDL code in a manner procedurally similar to that described above.  Strong absorption lines at $\lambda>1216$ \AA\ were flagged as possible intervening \lya\ lines.  Candidate \lya\ lines were confirmed by checking for counterpart \lyb\ and \lyg\ absorption before being accepted.  Equivalent widths were measured interactively from the normalized \lya\ profiles.  As above, multiple components were not used unless there was compelling evidence for their existence in weaker Lyman lines.  The STIS/G140M data for Mrk\,876 were analyzed similarly.  In all, our analysis yielded consistent results with published data in terms of velocities and equivalent widths.

The distribution of \HI\ absorbers as a function of column density is a power law with a negative slope as discussed below \citep{Penton2,Penton4}.  Thus, the absorber list features many weak absorbers and relatively few strong systems.  The ratio of equivalent widths for unsaturated \HI\ absorbers is equal to the ratio of $f\lambda^2$ between the lines, a factor of 6.24 for \lya\ and \lyb.  Since the $3\sigma$ limiting equivalent width of good \FUSE\ data is typically 10-20~m\AA, we set an equivalent width threshold of $W_{\rm Ly\alpha}>80$~m\AA\ to have any realistic chance of detecting weaker Lyman series or ionic lines.  Profile fits to weak \lya\ lines typically give more accurate column densities than strong lines (in the absense of a sufficiently optically thin higher order line), so curve-of-growth (COG) fitting from multiple transitions is less crucial.

We also limited the sample to those absorbers at $z\leq0.3$ where the Lyman limit is still within the \FUSE\ band.  This limits our analysis to the ``local'', low-redshift universe.  Furthermore, as in \citet{Penton1}, any absorber within $cz=1500$~\kms\ of the AGN emission velocity was discarded as being potentially related to the AGN.  However, quasar outflow velocities have been measured at much higher velocities \citep{Crenshaw99,Kriss02} and absorbers within $\sim$5000 \kms\ of $v_{\rm AGN}$ were treated individually.  Broad, asymetric line profiles, especially in metal lines, were used as a criterion in a few cases to disqualify absorbers as intrinsic to the AGN system (N. Arav and J. Gabel, priv comm.).

Our final census is 171 known \HI\ absorbers with $z\leq0.30$ and $W_{\rm Ly\alpha}>80$ m\AA.  Of these, there are 148 absorbers at $z\leq0.21$, where \CIII\ $\lambda977$ absorption appears within the \FUSE\ range, and 129 potential \OVI\ $\lambda1032$ absorbers at $z\leq0.15$.  These absorbers, along with reference information, are listed in Table~2 and assigned identification numbers based on sight-line and redshift ordering.  Four sight lines in Table~1 (NGC\,985, Ton\,951, Mrk\,279, and Mrk\,290) are devoid of strong \lya\ absorbers ($W>80$ m\AA) and not listed in Table~2.  These sight lines were not analyzed for high Lyman series or metal absorption, but they contribute to the total redshift path length and are valid statistical contributors to the sample.

\section{Data Analysis}

Once the \lya\ absorbers were established with reliable redshifts and equivalent widths (either from the literature or from our own analysis of the archival data), we analyzed the \FUSE\ data for each absorber in up to six Lyman-series transitions: \lyb\ $\lambda$1025.722, \lyg\ $\lambda$972.537, \lyd\ $\lambda$949.743, \lye\ $\lambda$937.803, \lyz\ $\lambda$930.748, and \lyeta\ $\lambda$926.226.  We also analyzed the metal transitions \OVI\ $\lambda$1031.9261, \OVI\ $\lambda$1037.6167, and \CIII\ $\lambda$977.0201.

Data were retrieved from the archive and reduced locally using {\sc calfuse}v2.4\footnote{More detailed calibration information is available at {\tt http://fuse.pha.jhu.edu/analysis/calfuse\_intro.html}}.  Raw exposures within a single \FUSE\ observation were coadded by channel midway through the pipeline.  This can produce a significant improvement in data quality for faint sources (such as AGN) since the combined pixel file has higher S/N than the individual exposures and consequently the extraction apertures are more likely to be placed correctly.  Combining exposures also speeds up reduction time dramatically.  Reduced data were then shifted and coadded by observation to generate a final spectrum from each of the eight data channels.  The data were binned by three pixels; \FUSE\ resolution is typically 8-10 pixels or roughly 3 bins.

The data were normalized in 10 \AA\ segments centered on the location of each IGM absorber as follows: the locations of prominent Galactic ISM lines, IGM lines, and intrinsic absorption lines from the AGN were marked automatically.  Line-free regions were then selected interactively and fitted using Legendre polynomials of order less than six.  The signal to noise per binned pixel $(S/N)_{\rm pix}$ was established locally as the $1\sigma$ deviation from the mean in the line-free continuum regions within 5 \AA\ of the IGM line.  Given that a \FUSE\ resolution element is typically 8--10 raw pixels ($\sim$3 binned pixels) or $\Delta V \approx 20$ \kms, we adopt the relationship $(S/N)_{\rm res}=\sqrt{3}\,(S/N)_{\rm pix}$ throughout this study.  Absolute velocity calibration in \FUSE\ data is not determined to much better than a resolution element.  We calibrated each segment by setting Galactic \HH\ lines to $v_{\rm lsr}=0$.  If \HH\ lines were unavailable, low-ionization ISM lines (\ion{Ar}{1}, \ion{Si}{2}, \ion{Fe}{2}, \ion{P}{2}) in the area were used or, if necessary, other IGM lines.  This process was carried out for each transition of each absorber in each of up to four detector channels covering the wavelength region.  

Once a collection of normalized spectral segments was generated, the quantitative analysis began.  Each transition was examined and measured interactively in two ways.  A multi-component Voigt profile fit was performed on the IGM absorber as well as on any other nearby lines with free parameters, $v$, $b$, and $N_{\rm fit}$.  From $b$ and $N_{\rm fit}$, the equivalent width $W_{\rm fit}$ was determined.  We used as few absorption components as possible in our fits since arbitrary additional components always improve the fit visually.  Second, an apparent column density $N_{\rm a}$, line width, $W_{\rm a}$, and velocity centroid were determined via techniques described in \citet{SavageSembach91}.  Roughly half of the lines in our survey were fit using both apparent column and Voigt fitting techniques and the resulting column densities, velocity centroids, $b$-values, and equivalent widths are within the uncertainties in most cases for low or moderate-saturation lines.  Voigt fitting tends to give more reliable results for saturated lines and lines where blending and multiple components are present.  The apparent column method is superior in noisy data or in weak absorbers where a profile fit will be heavily influenced by noise.  All equivalent widths were shifted to the rest frame: $W_{\rm r}=W_{\rm obs}(1+z)^{-1}$.

\begin{figure} 
  \epsscale{1.2}\plotone{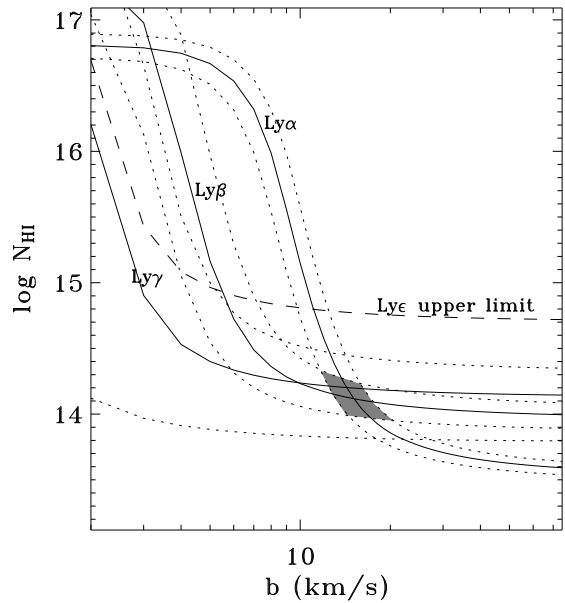} 
  \caption{Concordance plot for the absorber at $z=0.0071$ toward
  PKS\,1211+143 used to determine an accurate values of $b_{\rm HI}$
  and \NHI.  Any given equivalent width corresponds to a curve in
  $b$-$N$ space.  Multiple curves give accurate values for $N$ and
  $b$.  Dotted lines represent $1\sigma$ errors in measured
  $W_\lambda$ and the dashed line shows the upper limit measurement of
  $W_{\rm Ly\epsilon}$.  The Ly$\delta$ line for this absorber was
  blended with an \HH\ line and was unavailable.  The shaded area shows
  the derived $b$, $N$ value and associated uncertainty.}
\end{figure}

A 3$\sigma$ upper limit on equivalent width was also determined for each absorber based on the spectrograph resolution $R=\lambda/\Delta\lambda$ and local signal-to-noise per resolution element $(S/N)_{\rm res}$, \begin{equation} W_{\rm min}=\frac{3~\lambda}{R~(S/N)_{\rm res}}, \end{equation} where $R=\lambda/\Delta\lambda\approx15,000$.  In cases where no absorption was detected or when a detection was below a 3$\sigma$ level, the upper limit equivalent width was substituted.  In cases where a line was strongly blended, the total, blended equivalent width was used as an upper limit.

Roughly half the sight lines featured moderate to strong absorption by Galactic \HH\ at $\lambda<1120$~\AA\ out of rotational levels $J\leq4$.  We modeled and removed these lines when possible using a database of \HH\ column measurements compiled by \citet{Gillmon05}.  However, the modeled \HH\ profiles were not always precise enough to leave a flat, residual-free continuum in unblended \HH\ lines, so cases of \HH\ contamination in IGM lines were treated with caution.  When in doubt, the \HH\ was not modeled, and the fit to the contaminated line was treated as an upper limit.

Once all lines were measured, data from all four instrument channels were compared side-by-side.  This allowed us to interactively reject spurious detections (features appearing in only one channel) and instrumental features, and to systematically verify the existence of weak absorbers.  Of particular note is the strong instrumental ``absorption feature'' at 1043.5 \AA\ in the LiF1a channel \citep[see][for an overview of the \FUSE\ detectors and specifications]{Sahnow00}.  A surprising number of absorbers fell on or near this feature, and we were forced to use the lower-throughput LiF2b channel instead.  Fitted quantities (or upper limits) from the channel with the best quality data were chosen as representative of the line.  In the majority of cases, the LiF1 channel was used, but LiF2 was superior $\sim30\%$ of the time.  The SiC channels have much lower throughput than the LiF channels and were only used for lines that fell outside the LiF coverage ($\rm\lambda<1000~\AA$) or in the LiF chip gap ($\rm1084~\AA<\lambda<1087~\AA$).

\subsection{H\,I Analysis}

Column densities and $b$-values for the \HI\ absorbers were determined via COG concordance plots \citep[see][]{Shull00}, an example of which is shown in Figure~1.  Each value of equivalent width for a particular set of atomic parameters traces out a curve in $b$, $N$ space.  By plotting curves from several transitions with a high contrast in $f\lambda$, we can determine the accurate column density $N_{\rm cog}$ and $b$-value.  This method is equivalent to a traditional curve of growth fit, but it gives a better idea of the degree of uncertainty involved in both parameters.  In several cases, anomalously strong absorption in one Lyman line or another was caught and corrected by this technique.

Several \lya\ absorbers in our sample show clear evidence of multiple components, and many others undoubtedly feature unresolved structure.  Since our study focusses primarily on \OVI\ absorption, we have combined resolved \HI\ components into single absorbers in cases where only one, broad \OVI\ absorber is seen; for example, the pair of \HI\ absorbers at $z=0.0948$ and $0.0950$ toward PKS\,1302-102 show a single, broad \OVI\ line and thus have been combined into one absorber.  Other cases, where both \lya\ and \OVI\ show clear multiple structure, have been treated as separate absorbers.

To test our concordance plot method of determining $N$ and $b$, we simulated multi-component absorbers with a grid of $\sim100$ models varying the column density ratio ($N_1/N_2=1-10$) and velocity separation ($\Delta v=0-75$ \kms).  We measured the equivalent width of the blended pair of simulated absorbers and treated them as a single absorber without any a priori knowledge of their structure.  For a pair of simulated \HI\ absorbers with column densities $N_1$ and $N_2$ and line widths $b_1$ and $b_2$ separated by $\Delta v$, a COG analysis of the blended system yields a combined column density no more than 20\% greater than $N_1+N_2$ and usually substantially better.  The derived line width {\em is} affected by unresolved structure as component separation mimics a larger $b$ value.  Nevertheless, we find empirically that $b_{\rm COG}^2\leq b_{\rm max}^2+0.45\,(\Delta v)^2$ where $b_{\rm max}$ is the larger of the two component linewidths.  Multiple components, for instance multiple absorbers within the same galaxy cluster, may mimic a broader single absorber, but the effect is small for line separations of \FUSE\ instrumental resolution or less.  Total column density is essentially insensitive to blended sub-components via our methods.  This result is consistent with that found by \citet{Jenkins86}.

Our COG-derived $N$ and $b$ values for the \HI\ absorbers, along with $1\sigma$ limits, are listed in Table~3.  In some cases, particularly for weak absorbers, concordance plots yielded no useful information and we adopt $N$ and $b$ from other sources.  In most of these cases, we adopt measurements from \lya-only fits or apparent column measurements quoted in the literature or measured by the Colorado group.  In a few cases, we assume $b=20\pm10$ \kms\ and derive $N$ based on the observed \lya\ equivalent width.  These cases are noted in Table~3.

As a further check on the validity of our methods, we checked our $b_{\rm HI}$ and $N_{\rm HI}$ values against published values.  There are surprisingly few COG-qualified \HI\ measurements in the literature and we were only able to find 30 \HI\ absorbers for our comparison.  Our \NHI\ values are consistent within $\pm1\sigma$ with published values in 20 out of 30 cases (67\%), within $\pm2\sigma$ in 24 cases (80\%), and within $\pm3\sigma$ in all cases.  Doppler width is consistent within $\pm1\sigma$ in 18 out of 28 cases (64\%, two absorbers were disqualified from consideration because they were interpretted as multiple absorbers in the litterature but as a single absorber in this study) and within $\pm2\sigma$ in 26 cases (93\%).  We have noted $>2\sigma$ deviations from published values in Table~3.  The median absolute deviance in the sample is $\sim0.6\sigma$ for both $b\rm_{HI}$ and \NHI.  The two-sided deviance distribution in each case is symetric and consistent with no systematic over- or under-prediction.  

\subsection{O\,VI and C\,III Analysis}

The \OVI\ doublet does not have enough contrast in $f\lambda$ to use concordance plots to determine accurate columns, nor are both lines detected for many absorbers.  Apparent and/or profile-fit column densities and $b$-values were adopted for these transitions.  The equivalent width of absorption lines on the linear part of the curve of growth follows the relation 
\begin{equation}W_\lambda=\biggl(\frac{\pi e^2}{m_e c}\biggr)\frac{N f \lambda^2}{c}\end{equation}
while line-center optical depth goes as 
\begin{equation}\tau_0=\biggl(\frac{\pi e^2}{m_e c}\biggr)\frac{N f \lambda}{\sqrt{\pi}\,b}.\end{equation}
Thus an absorbsion line saturates (reaches $\tau_0=1$) at equivalent width $W_\lambda{\rm(sat)}=(b/c)\sqrt{\pi}\lambda=(153~{\rm m\AA})\,b_{25}$ where $b_{25}$ is the Doppler parameter in units of 25 \kms.  This corresponds to $N_{\rm HI}=(10^{13.62}$ cm$^{-2})\,b_{25}$ for \lya.  For \OVI, saturation occurs at a column density \NOVI\ $\sim(10^{14.09}$ cm$^{-2})\,b_{25}$ and most of our \OVI\ detections are at or below this level (see Figure~1a of Paper~I).  Similarly, \CIII\ saturates at \NCIII\ $\sim (10^{13.35}$ cm$^{-2})\,b_{25}$ which is within the range of our measurements.  Any saturation in ionic absorbers is mild at worst, and Voigt profile fits and apparent column measurements are probably indicative of the true values.  For upper limit cases, the S/N-based $3\sigma$ minimum equivalent width was converted to column density via a curve of growth assuming $b=25$~\kms.

Our measured line widths and equivalent widths for the \OVI\ doublet absorbers are listed in Table~4.  In cases where both lines of the \OVI\ doublet were detected, an error-weighted mean of the columns was taken as the canonical value.  Quoted errors are $1\sigma$ uncertainties based on line fits.  In some cases, absorption is seen at velocities different from the expected value.  Part of the difference is due to the variable \FUSE\ wavelength solution and uncertainty in the fiducial \lya\ absorber velocity, but $\Delta v>30$ \kms\ probably represents a real kinematic difference.  High-$\Delta v$ cases are noted in Table~4.

\CIII\ has only one resonance transition at 977.02 \AA\ in the \FUSE\ band, which makes detection more challenging than for the \OVI\ doublet lines.  Furthermore, low-redshift \CIII\ absorbers must be measured at $\lambda < 1000$~\AA\ where the \HH\ line density is high and \FUSE\ S/N is low.  On the other hand, the short rest wavelength of \CIII\ allows absorbers to be measured out to a higher redshift ($z\leq0.21$) in \FUSE\ data.  Column densities and $b$-values for \CIII\ were determined in the same way as \OVI\ and are listed in Table~5.  In several cases, multiple \CIII\ components were seen in an evidently single \HI\ absorber.  These were measured separately, and the quantities listed in Table~5 represent the total equivalent width and column for the system.

\begin{figure*} 
 \plottwo{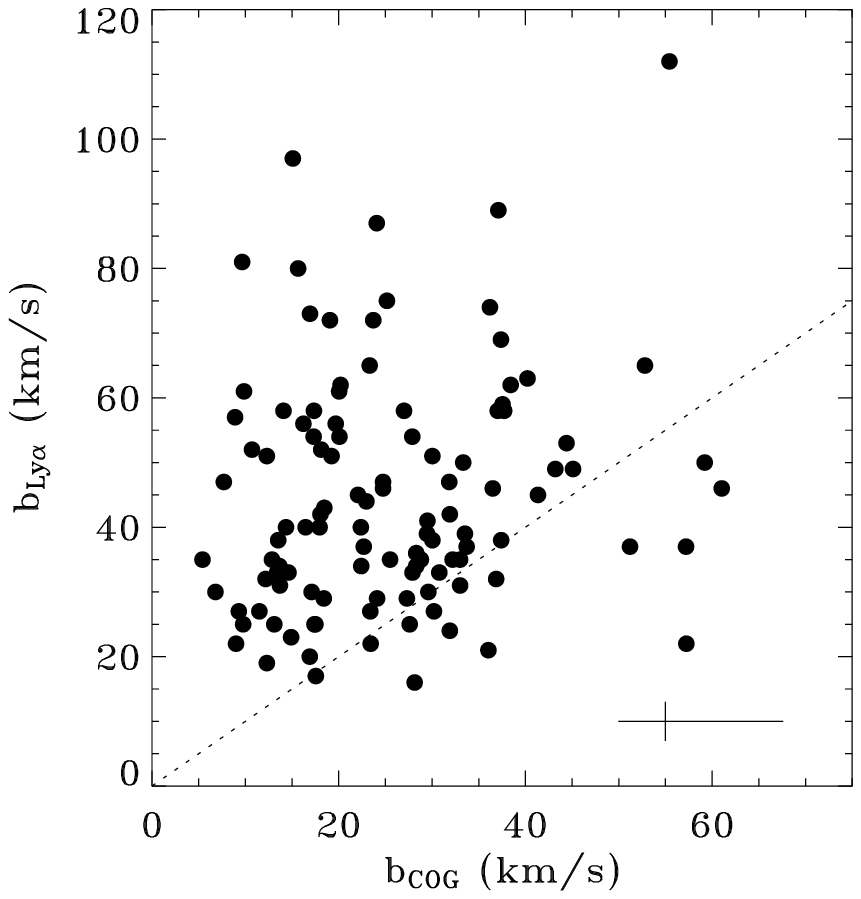}{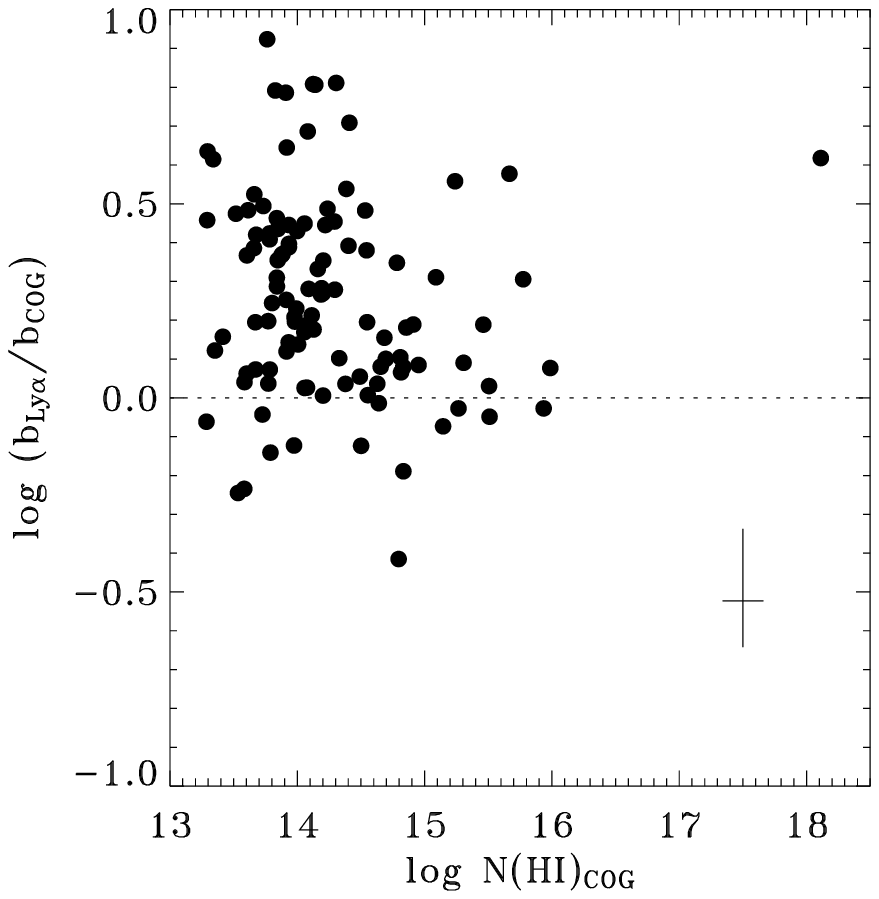} 
 \caption{(Left) Comparison of line width measured from \lya\ lines
 alone versus curve-of-growth $b$ from our concordance plots.  (Right)
 $b$ value predictive accuracy as a function of \NHI.  There is little
 correlation between the $b$ measured from concordance plots ($b_{\rm
 COG}$) and from the \lya\ line alone ($b_{\rm Ly\alpha}$), but
 $b_{\rm Ly\alpha}$ tends to overpredict $b_{\rm COG}$ values.  The
 ratio $b_{\rm Ly\alpha}/b_{\rm COG}$ has a faint dependence on column
 density, with the widths of stronger absorbers being more accurately
 predicted by \lya-only measurements in comparison to lower \NHI\
 lines.  Median uncertainties are shown in the lower right of each
 panel.}  \label{fig:bvsb}
\end{figure*}

As a final step in our analysis, we carefully inspected each candidate \OVI\ and \CIII\ absorber by hand to verify its existence.  Consequently, some detections were downgraded to upper limits because of blending or suspect data features such as flat-fielding issues.  In total, we detect 30 \CIII\ absorbers and 88 upper limits.  \OVI\ is detected in one or both lines of the doublet in 40 absorbers with 84 upper limits.


\section{Discussion}

\subsection{The Importance of \lyb}


The majority of the literature on IGM absorption lines, particularly at high redshift, is based on analysis of \lya\ lines.  A few words of caution regarding this are in order.  Our curve-of-growth measurements of \NHI\ and $b_{\rm HI}$ using multiple Lyman lines for each absorber should be more accurate than measurements based on \lya\ detections alone; \lya\ lines saturate at log\,$N_{\rm HI}\sim13.5$ ($W_{\rm Ly\alpha}\sim180\,b_{25}$~m\AA).  We expect \lya-only measurements to accurately describe the column and width of weak absorbers where saturation is minimal, but the \lya-only measurements should grow increasingly inaccurate as saturation and unresolved subcomponents become more substantial.  The median $W_{\rm Ly\alpha}$ in our sample is greater than 200~m\AA, so \lya\ saturation is a definite concern.

Figure~\ref{fig:bvsb}a shows no correlation between COG-determined Doppler width, $b_{\rm COG}$, and line-width based on the \lya\ absorbers alone, $b_{\rm Ly\alpha}$.  In general, $b_{\rm Ly\alpha}$ overpredicts the multi-line $b_{\rm COG}$ by a factor of two or more as seen in \citet{Shull00}, but there is no other correlation.  Using high-resolution data, \citet{Sembach01} found multiple narrow velocity components within the \lya\ absorption complex at $z=0.0053$.  Their COG analysis of this system using \lya-Ly$\theta$ lines shows $b=16.1\pm1.1$ \kms\ and log\,$N_{\rm HI}=15.85^{+0.10}_{-0.08}$ while \lya-only measurements give $b=34.2\pm3.3$ \kms\ and log\,$N_{\rm HI}=14.22\pm0.07$, a factor of two in line width and 43 in column density.  Figure~\ref{fig:bvsb}b shows $b_{\rm Ly\alpha}/b_{\rm COG}$ as a function of \NHI; the stronger absorbers show less line-width overprediction than the weaker absorbers.  However, given the heterogenous nature of the methods used to measure \lya\ line widths and column densities, we are hesitant to spend too much effort analyzing their differences.  

\begin{figure} 
  \epsscale{1.2}\plotone{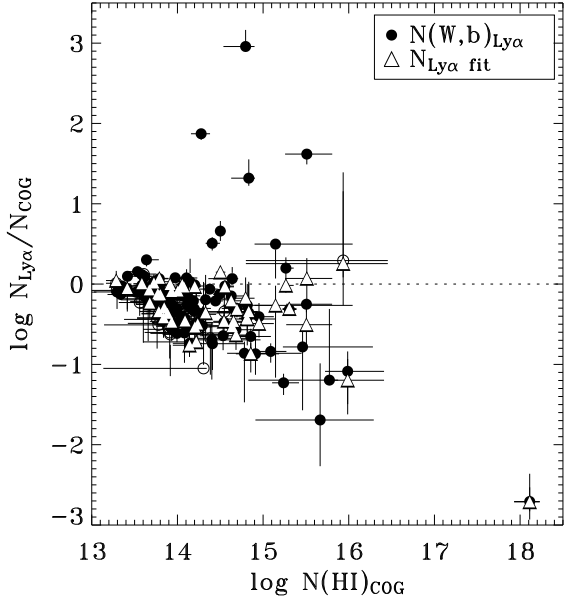} 
  \caption{A comparison of column density determined from \lya\ lines
  alone vs curve-of-growth column density.  \NHI\ determined from
  $W_{\rm Ly\alpha}$ and \lya\ line width (circles) is less than that
  determined from a COG using multiple Lyman lines.  The
  underprediction generally becomes worse for more saturated \lya\
  absorbers.  A value $b=25$~\kms\ is assumed for any absorbers with
  no listed \lya\ linewidth (open circles).  Fitting \lya\ lines with
  Voigt profiles (triangles) generally gives a better match, but still
  tends to underpredict \NHI\ from a COG.} \label{fig:nvsnh1}
\end{figure}

The differences between curve-of-growth column densities and \lya-only column densities can be addressed more rigorously.  The literature sources listed in Table~2 measure \lya\ column density in a variety of ways.  We ignore the quoted column density and calculate \NHI\ based only on $W_{\rm Ly\alpha}$ using a single-transition concordance plot and the quoted $b$ value.  For absorbers where no line width is quoted, we choose $b=25$ \kms\ based on our COG mean value as discussed below.  This gives us a consistent, \lya-only based column density to compare with our multi-line curve of growth results.  Figure~\ref{fig:nvsnh1} shows that \lya\ is reasonably accurate in predicting the column densities of weak, unsaturated lines (log\,$N\la13.5$).  However, \NHI\ is increasingly underpredicted for stronger, more saturated lines.

A subset of literature sources, including all of the absorbers measured by the Colorado group, determined \lya-only column densities from profile fits in which \NHI\ and $b$ were free parameters.  These are shown in Figure~\ref{fig:nvsnh1} as open triangles.  We find that a similar trend is present even in the profile-fit data, although it tends to be more accurate than columns based on $b_{\rm Ly\alpha}$ and $W_{\rm Ly\alpha}$.  These points serve to further illustrate the critical importance of higher Lyman lines to the analysis of \HI\ absorbers in the IGM.


The number of absorbers $\cal N$ as a function of column density \NHI\ follows a power-law distribution $d{\cal N}_{\rm HI}/dN_{\rm HI} \propto N_{\rm HI}^{-\beta}$ so that the integrated number distribution ${\cal N}(\ge N_{\rm HI})\propto N_{\rm HI}^{1-\beta}$.  \citet{Penton4} obtained values of $\beta=1.65\pm0.07$ over the column density range $12.3 \leq {\rm log}~N_{\rm HI} \leq 14.5$ and $\beta=1.33\pm0.30$ for $14.5 \leq {\rm log}~N_{\rm HI} \leq 17.5$, based on a similar set of absorbers and an assumed $b=25$ \kms.  \citet{Williger06} find $\beta=2.06\pm0.14$ for log\,\NHI$\geq13.3$ based on 60 \HI\ absorbers.  Our sample is limited to $W_{\rm Ly\alpha}>80$ m\AA, so much of the lower range in column density is missing when compared to the Penton sample, but we recreate the power-law fits to the distribution using $b$ and \NHI\ values from curves of growth.  We find $\beta=1.85\pm0.39$ for the weak absorbers ($13.8 \leq {\rm log}\,N_{\rm HI} \leq 14.5$) and $\beta=1.50\pm0.23$ for the stronger absorbers ($14.5\leq {\rm log}\,N_{\rm HI} \leq 17.5$) but note that the individual sub-samples are small and may be of reduced statistical significance.  This slight steepening of the distribution between \lya-only and full COG analyses is contrary to what we would expect.  Given that \lya-only measurements tend to underpredict column density relative to full COG analyses as shown in Figure~\ref{fig:nvsnh1}, we would expect a flatter distribution, especially for stronger absorbers.

\begin{figure}  
 \epsscale{1.2}\plotone{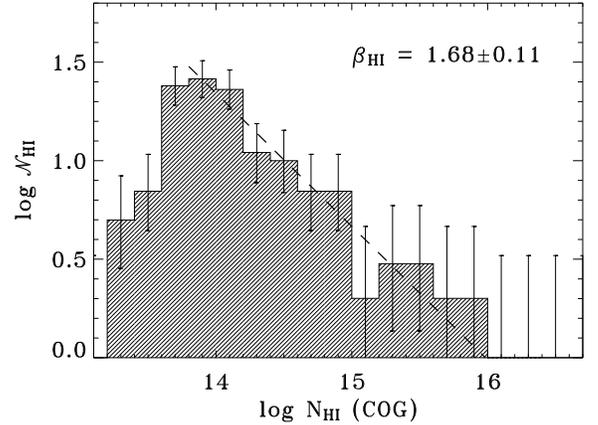} 
 \caption{The revised distribution of HI absorbers as a function of
 column density; error bars represent single-sided counting errors
 \citep{Gehrels86}.  The distribution can be fitted by a power law
 $d{\cal N}_{\rm HI}/dN_{\rm HI}\propto N_{\rm HI}^{-\beta}$ so that
 ${\cal N}(\ge N_{\rm HI})\propto N_{\rm HI}^{1-\beta}$.  We find
 $\beta=1.68\pm0.11$ for our sample above $N_{\rm HI}=10^{13.8}$
 cm$^{-2}$ (dashed line).} \label{fig:nh1powerlaw} 
\end{figure}

Simulations \citep{Dave01} suggest that higher column density absorbers will evolve faster toward lower redshift than low column absorbers through infall and large scale structure formation, thus producing a steeper distribution at $z=0$.  \citet{Weymann98} find $\beta\approx1.3$ for log\,$N_{\rm HI}>13.5$ assuming $b=30$ \kms\ for sight lines in the FOS Key Project ($z<1.3$).  We find a considerably steeper distribution in our total sample as shown in Figure~\ref{fig:nh1powerlaw}: $\beta=1.68\pm 0.11$ for log\,$N_{\rm HI}>13.8$.  Studies of the high-redshift \lya\ forest find $\beta=1.46$ and $1.55$ at $z\sim2.85$ and $z\sim3.7$, respectively \citep{Hu95,Lu96,Kim97,DaveTripp01}.  Our distribution is marginally steeper than the high-redshift samples, but does not agree with the low-redshift FOS sample.  The FOS survey is limited to \lya-only measurements of strong absorbers ($W_{\rm Ly\alpha}\ga200$ m\AA) and closely-spaced absorbers individually near the detection threshhold may be blended together into an artificially strong single absorber \citep[see e.g.][]{ParnellCarswell88}.  This would artificially flatten the FOS distribution with respect to more sensitive, higher resolution surveys.

\begin{figure*} 
 \plottwo{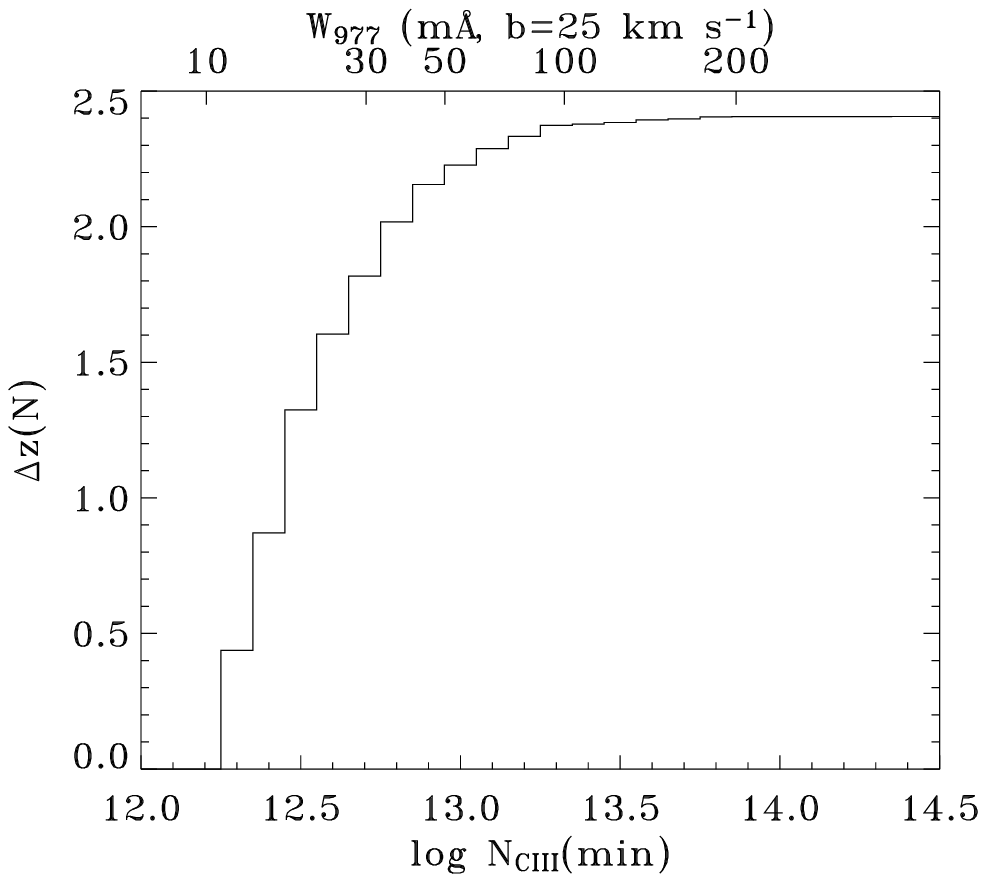}{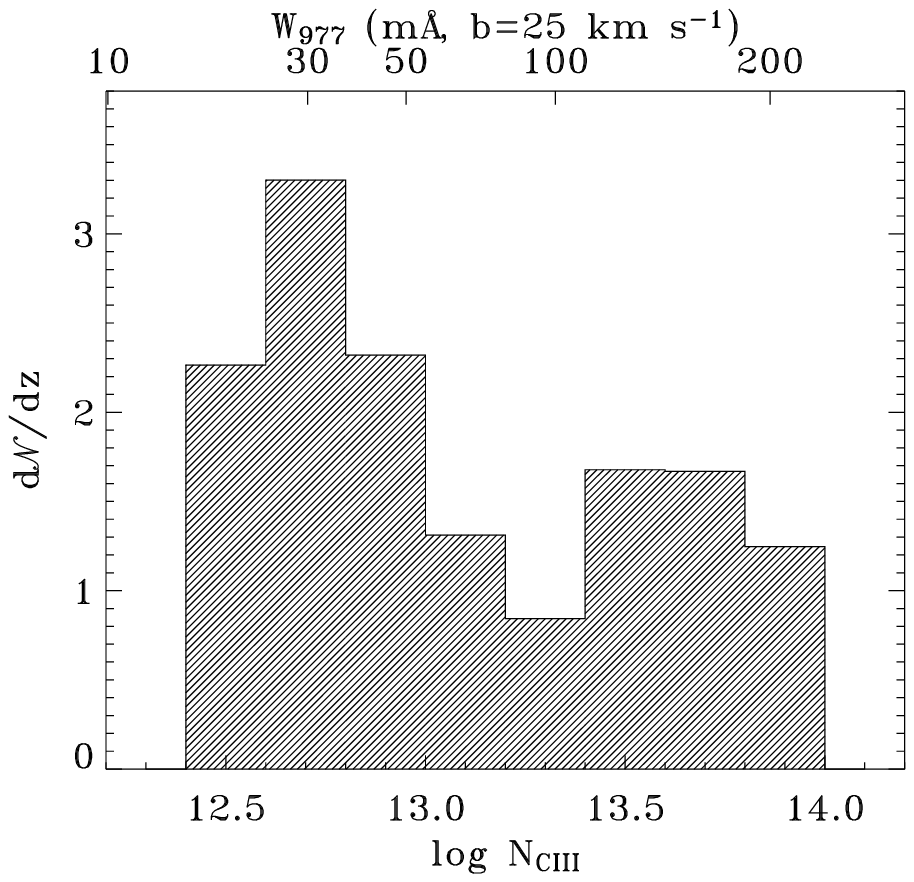}
 \caption{The profile of effective redshift versus column density
 sensitivity $\Delta z(N)$ for C\,III (left) is used to correct for
 sample incompleteness in $d{\cal N}_{\rm CIII}/dz$ as a function of
 column density in cm$^{-2}$ (right).  The equivalent-width limit for
 the $\lambda977$ line is shown at the top of both figures.}
 \label{fig:c3plot}
\end{figure*}

\subsection{C\,III Absorber Distribution}

The \CIII\ detections are not as numerous as \OVI\ detections.  There is only one resonant \CIII\ transition, and each \CIII\ absorber stands a higher chance of being obscured by another line, often \HH.  The \CIII\ \lam977 transition is intrinsically 5.4 times stronger (in $f \lambda$) than \OVI\ \lam1032.  If intergalactic C/O has a relative abundance ratio equal to that in the Sun, (C/O)$_{\odot} = 0.5$ \citep{Allende01}, the C~III transition would be 2.7 times more sensitive to metal-enriched gas than O~VI.  Counteracting the intrinsic \CIII\ strength is the fact that low-$z$ C~III absorbers fall in the less sensitive (SiC) portion of the \FUSE\ detectors.

We see \CIII\ at the $>3\sigma$ level in 30 out of 148 \lya\ absorbers.  To calculate the total redshift pathlength, we employ a method identical to that used for \OVI\ in Paper~I.  The $3\sigma$ minimum equivalent is calculated as a function of wavelength based on signal to noise for each sight line.  This minimum equivalent width is translated into $N_{\rm min}(\lambda)$ assuming $b=25$ \kms.  An absorber is then moved along the spectrum as a function of $z_{\rm abs}$ and the total pathlength is summed as a function of $N_{\rm min}$ for \CIII\ (Figure~\ref{fig:c3plot}a).  Detections are binned by column density, and the bins are scaled by $\Delta z(N)$ to produce $d{\cal N}_{\rm CIII}/dz$ (Figure~\ref{fig:c3plot}b).  We define the rolloff in the redshift pathlength as the point where $\Delta z(N)$ equals 80\% of $\Delta z(\infty)$.  This occurs at log\,\NCIII\ = 12.75, compared to log\,\NOVI\ = 13.35 (see Figure~3, Paper~I).  

There is little if any literature information on \CIII\ absorber statistics.  We quote here the integrated line frequency per unit redshift, $d{\cal N}_{\rm CIII}/dz$, down to a given equivalent width limit, in parallel with $d{\cal N}_{\rm OVI}/dz$ (Paper~I and sources therein).  Above $W_{\lambda} = 30$~m\AA, we see 26 \CIII\ absorbers over $\Delta z\le2.406$, so that $d{\cal N}_{\rm CIII}/dz = 12^{+3}_{-2}$.  The quoted uncertainties are based on the single-sided $1\sigma$ confidence limits of Poisson statistics \citep{Gehrels86}.  Raising the threshold to $W_{\lambda} > 50$ \AA\ yields 20 \CIII\ absorbers and $d{\cal N}_{\rm CIII}/dz = 8\pm2$.  We fit the cumulative distribution in column density assuming a differential power law, $d{\cal N}_{\rm CIII}/dN_{\rm CIII} \propto N_{\rm CIII}^{-\beta}$ as we have with \OVI\ (Paper~I) and \HI\ \citep[this work]{Penton4}.  We find $\beta_{\rm CIII}=1.68\pm0.04$ for $12.4<{\rm log}\,N_{\rm CIII}<13.8$, remarkably similar to $\beta_{\rm HI}$ but not as steep as the \OVI\ distribution.  The similarity in $\beta_{\rm HI}$ and $\beta_{\rm CIII}$ is circumstantial evidence that both species are tied to related ionization processes, while the steeper $\beta_{\rm OVI}$ suggests that a different physics is taking place here.  We explore the issue of ionizing mechanisms below.

\subsection{The Multiphase IGM}

A central question in the IGM, and one critical to our interpretation of the data, is whether the observed gas is collisionally ionized through shocks and conductive interfaces, photoionized by the ambient QSO radiation field, or some combination of the two.  We have usually assumed that \OVI\ is created in shocks between intergalactic clouds (Paper~I) and thus traces the WHIM phase.  However, an ambient hard UV field incident on very low density gas ($n_H\sim10^{-5}$ cm$^{-3}$) can produce a measurable quantity of \OVI\ and other high ions.  If this is the case, the observed \OVI\ does not represent a hot phase at all and cannot be used to trace the WHIM.

\citet{Dave01} argue that the radiation field at $z>2.5$ was strong enough to ionize a significant amount of the oxygen in the IGM to \OVI, but that the ionizing flux at $z<1$ is too low; QSOs are less common in the modern universe, and the Hubble expansion has spread them to greater average distances diluting the radiation field.  Shocks are therefore the likely source of highly ionized gas, either from SN-driven galactic winds, virial shocks associated with large galaxies and clusters, or shocks from infalling clouds outside the virial radius \citep{BirnboimDekel02,Shull03,Furlanetto05,Stocke05}.

On the other hand, \citet{Prochaska04} analyzed metal lines in six absorbers along the sight line toward PKS\,0405-123 on the basis of collisional and photoionization models.  They determine that several absorbers are consistent with single-phase collisional ionization under collisional ionization equilibrium (CIE), but that others are better explained with by photoionization from an ambient QSO field.  They note, however, that several systems could also be explained with multi-phase ionization models.  Similarly, \citet{Collins04} analyze absorption lines from many different FUV ions in four high velocity cloud absorbers in the Galactic halo.  They conclude that some absorbers show evidence for photoionization while others are more consistent with ionization by shocks.  In at least one case, the best explanation was a combination of photo- and collisional ionization.  The question of shocks versus photoionization is not yet resolved \citep{Stocke05}.

Our study does not include enough different ion states for a detailed comparison of each absorber with models.  Future work on \ion{C}{4}, \ion{Si}{4}, and \ion{Si}{3} absorption in STIS data should constrain observations so that detailed ionization models in can be constructed.  However, we can look at \HI, \CIII, and \OVI\ absorber detections to examine the energetics of each set as a specific class as well as the ensemble as a whole.

The absorber detection statistics are instructive.  For the purposes of this argument, we define detections as \HI\ absorbers with log\,\NOVI$\ge13.2$ (37 out of the 40 total $3\sigma$ detections) or log\,\NCIII$\ge 12.6$ (27 out of 30 total detections).  Similarly, we define non-detections as upper limits below the thresholds.  This accounts for the top $\sim90$\% of detections in each metal ion and counts only upper limits with reasonable data quality.  We further narrow the sample to those absorbers with good statistics (detections or non-detections as defined above) in both \OVI\ and \CIII.  This gives a sample of 45 \HI\ absorbers with absorption statistics in both metal ions.  We detect \OVI\ in 20 absorbers (44\%) and \CIII\ in 16 (36\%), and twelve absorbers (27\%) show both \OVI\ and \CIII.  We see \HI\ and \OVI\ without accompanying \CIII\ in eight cases (18\%), and four absorbers (9\%) show \HI\ and \CIII\ without \OVI.  Twenty-one absorbers show neither \OVI\ nor \CIII.

An important caveat is that our survey for metal absorbers is not a ``blind'' survey; we look for absorption only near the velocities of known, strong \lya\ absorption, and we ignore the possibility of low-ionization metal-line absorbers without \HI.  Except for highly ionized (X-ray) absorption lines, there is no definitive evidence for such absorption in the low-redshift universe.  Thus, we treat $\Delta z_{\rm metal}$ as equivalent to $\Delta z_{\rm HI}$.  Furthermore, we assume that our sample of IGM absorbers have \HI, \CIII, and/or \OVI\ which are cospatial but not necessarily well-mixed; we limit velocity separations between any two lines to $\Delta v<100$ \kms.  We do not have the spectral resolution for detailed kinematic studies of the absorption systems.


\subsubsection{Absorber Line Widths}

\begin{figure} 
 \epsscale{1.2}\plotone{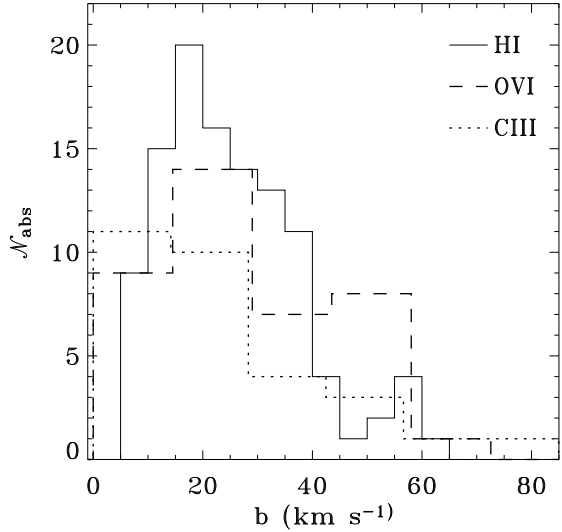} 
 \caption{Distribution of $b$ values for H\,I, O\,VI, and C\,III
 absorbers.  The $b\rm_{COG}(HI)$ values represent COG values while
 $b\rm_{cor}(metal)$ have been corrected for an instrumental
 resolution of $\Delta V \approx 15$ \kms\ ($b_{\rm FUSE}=9$ \kms).
 Mean values are $\langle b\rm_{COG}(HI)\rangle=25\pm13$ \kms,
 $\langle b\rm_{cor}(OVI)\rangle=31\pm14$ \kms\ and $\langle
 b\rm_{cor}(CIII)\rangle=26\pm13$ \kms.}  \label{fig:bvalhist}
\end{figure}

The distribution of Doppler line widths is a clue to the temperature (and hence phase) of a species and may help clarify the ionization source.  The $b$ values of \OVI\ and \CIII\ absorbers both show asymmetric distributions that peak around 20 \kms\ and fall off gradually to about 50 \kms.  Unlike the $b_{\rm COG}$ values derived for the \HI\ absorbers, $b\rm(metal)$ includes an instrumental profile.  We assume $R=15,000$ for a \FUSE\ resolution of 15 \kms\ \citep{Hebrard05,Williger05} ($b_{\rm FUSE}=12$~\kms) and subtract this in quadrature from $b_{\rm metal}$.  We find mean values $b_{\rm cor}(\rm OVI)=29\pm18$ \kms\ and $b_{\rm cor}(\rm CIII)=23\pm18$ \kms\ (Figure~\ref{fig:bvalhist}).  Median values are smaller by $\sim3$ \kms\ in both cases.

Thermal line widths for \OVI\ and \CIII\ at their peak CIE temperatures are $b\rm_{therm}\sim17$ \kms\ and $b\rm_{therm}\sim10$ \kms, respectively; the observed line widths are consistent with multiple photoionized components or thermal broadening of a collisionally ionized component.  Regardless of temperature, some of the observed width must be due to turbulence, multiple components, or Hubble expansion within extended, diffuse clouds \citep{Weinberg98,Richter04}.  \citet{Heckman02} argue that, since the fractional abundance of \OVI\ in CIE is a function of temperature peaked at $T_{\rm max}=10^{5.45}$~K, \OVI\ preferentially is found at or near its CIE peak temperature as gas cools from higher temperatures and $T\approx T_{\rm OVI}=10^{5.45}$~K is a good approximation.  (In non-equilibrium cooling, the gas could start out at $10^6$~K or hotter and cool below $T_{\rm max}$.) If $T\approx T_{\rm max}$, subtracting a thermal line width component at $T_{\rm OVI}$ leaves $b\rm_{turb}(OVI)=23$ \kms, equivalent to the Hubble broadening for a quiescent cloud $\sim330\,h_{70}^{-1}$ kpc across.  Of course, we do not know the extent of turbulent velocities in IGM clouds, so this should be considered an upper limit for the typical WHIM absorber scale (see discussion in \S~4.3.4 below).

Beryllium-like \CIII\ (1s$^2$2s$^2$) has two valence electrons and thus has a broader range of temperatures over which it has a significant ionization fraction in CIE.  Because little theoretical work has been done on \CIII\ ionization fraction in the non-equilibrium conditions likely to be present in the IGM, we assume that its peak CIE abundance temperature, $T_{\rm CIII}=10^{4.85}$~K, is characteristic of any collisionally ionized \CIII\ absorbers.  This corresponds to $b_{\rm thermal}(\rm CIII)=10$ \kms.  Photoionized \CIII\ would have a lower temperature ($\sim10^4$~K) and much smaller thermal correction of $b_{\rm therm}\sim4$ \kms.  Thus, $b\rm_{turb}(CIII)=21-23$ \kms, depending whether \CIII\ is collisionally ionized or photoionized, similar to $b\rm_{turb}(OVI)$.

Hydrogen line width is more sensitive to temperature than the metal ions.  The distribution of $b\rm_{COG}(HI)$ is peaked near 20 \kms, but it has non-negligible contribution from values out to 35 \kms\ implying contributions from turbulence, Hubble expansion, or unresolved multi-component structure.  The mean value is $25\pm13$ \kms.  Since $b\rm_{COG}(HI)$ is derived from equivalent widths, it is independent of instrumental resolution and represents only the quadratic sum of thermal and turbulent components.  If the entire $b\rm_{COG}(HI)$ is thermal in nature (which is unlikely), this represents gas with a temperature of log\,$T=4.6^{+0.4}_{-0.6}$ using $1\sigma$ errors to define a temperature range.  The thermal width of a line at typical \HI\ temperatures of $10^4$~K accounts for $b\rm_{therm}(HI)=13$ \kms\ leaving $b\rm_{turb}(HI)=21$ \kms, similar to that seen in \CIII\ and \OVI.

It may be that the observed \HI\ absorption comes not from a warm phase, but from the WHIM.  \citet{Richter04,Richter05} note that the neutral fraction of hydrogen at WHIM temperatures in CIE is very small ($<10^{-5}$).  However, a large total hydrogen column density (log\,$N_{\rm H}\sim19$) of low-density gas spread over 300 kpc will still produce measurable absorption in \HI.  These systems would be distinguishable from non-WHIM \HI\ absorbers by their large thermal $b$ values.  For purely thermal line-widths, we would expect $b_{\rm HI}\sim70$ \kms\ for $T=10^{5.5}$~K.  \citet{Richter05} report on 26 features in the sight lines of four AGN which they identify as broad \lya\ absorbers.  We see no absorbers with $b_{\rm HI}>70$ \kms, but there are 8 absorbers with $b_{\rm HI}>45$ \kms.  These may be WHIM \HI\ or unresolved multiple components as noted above.  However, the majority of the \HI\ lines show narrow line widths, inconsistent with hot \HI, and we are forced to conclude that \HI\ traces a warm neutral and moderately ionized phase in most cases.  The issue of gas temperature is still open for \CIII\ and \OVI, at least based on line widths.

\begin{figure} 
 \epsscale{1.2}\plotone{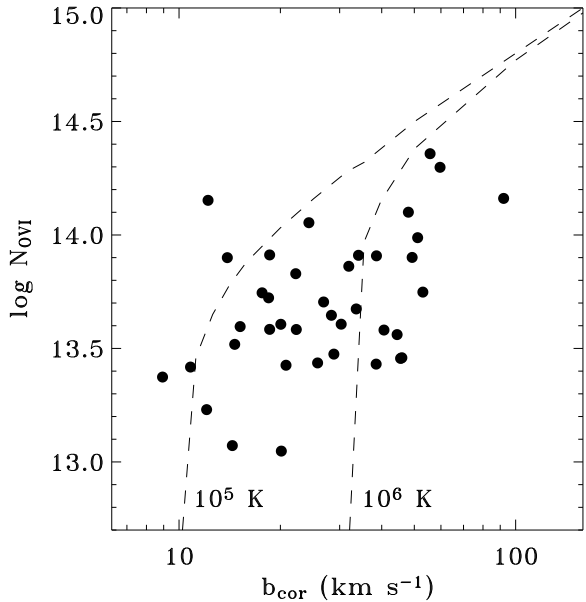} 
 \caption{Theoretical models show a relationship between $b_{\rm OVI}$
 and \NOVI\ (cm$^{-2}$) across a wide range of astrophyical systems.
 The dashed lines show the predictions of \citep{Heckman02} for
 temperatures of $10^5$~K and $10^6$~K.  We see no such correlation in
 IGM O\,VI absorbers in our 40 absorbers (solid circles), although
 they do tend to fall within the bounds of WHIM temperature curves.}
 \label{fig:heckmanNb}
\end{figure}

\citet{Heckman02} argue that line width is directly related to the velocity of the postshock gas and that all \OVI\ absorption systems, regardless of scale and origin, follow a simple relationship between line width and column density.  The exact relationship depends on the temperature of the \OVI\ gas, and comes about as hot gas cools radiatively through the temperature regime in which \OVI\ is abundant and can be detected.  Observational data from systems as diverse as starburst galaxies ($N_{\rm OVI}\sim10^{15}$ cm$^{-2}$) and Galactic disk absorbers ($N_{\rm OVI}\sim10^{13-14}$ cm$^{-2}$) show good agreement with theoretical predictions \citep[see][Figure~1]{Heckman02}.  The eight IGM data points in that Figure are located at the lower end of the column density range and show the most spread of any physical system.

With 40 \OVI\ detections in the IGM, we can improve the statistics immensely.  Figure~\ref{fig:heckmanNb} shows the relationship between corrected $b_{\rm cor}(\rm OVI)$ and \NOVI.  The IGM \OVI\ absorbers from our sample do not show the same correlation as absorbers in other physical situations, although the data points generally fall between the temperature curves corresponding to $10^5$~K and $10^6$~K.  The lack of correlation may be due to unresolved multiple components in the IGM absorbers or Hubble broadening.  However, blended absorbers would show a higher $b$ and be located farther to the right than they should.  It is clear that many of the observed IGM systems already show smaller $b$ than absorbers from different systems.

The explanation may be that \citet{Heckman02} rely on complete radiative cooling in their model to derive the $N-b$ correlation.  All the other systems in question have physical densities several orders of magnitude higher than expected for the IGM ($n_H\sim10^{-5}$ cm$^{-3}$) and thus the cooling time is comparatively short.  Diffuse IGM clouds can have cooling times comparable to or longer than $H_{\rm 0}^{-1}$ \citep{Collins04,Furlanetto05}.  Once a given species is out of ionization equilibrium, it becomes harder to use its line width as a gauge of temperature.  For example, a gas that cools faster than it recombines could have a narrower line width than derived under CIE assumptions.  


\subsubsection{Column Density Ratios}

In Paper~I we demonstrated a lack of correlation between \NOVI\ and \NHI.  \NHI\ varies over a range of nearly 1000 in our sample of low-column systems and the Lyman limit systems (LLSs).  Damped \lya\ absorbers (DLAs) seen in other sight lines \citep{Lopez99,Jenkins03} extend the range of \NHI\ over at least eight orders of magnitude.  In contrast, \NOVI\ shows a variation of only a factor of 30 in our sample and has never been detected at columns of greater than a $\rm few\times 10^{15}$~cm$^{-2}$ \citep{Heckman02}.  This small range of \NOVI\ compared with \NHI\ manifests itself as a good correlation of the ``multiphase ratio'' \NHI/\NOVI\ with \NHI\ (see Paper~I, Figure~1).

\begin{figure*} 
 \plottwo{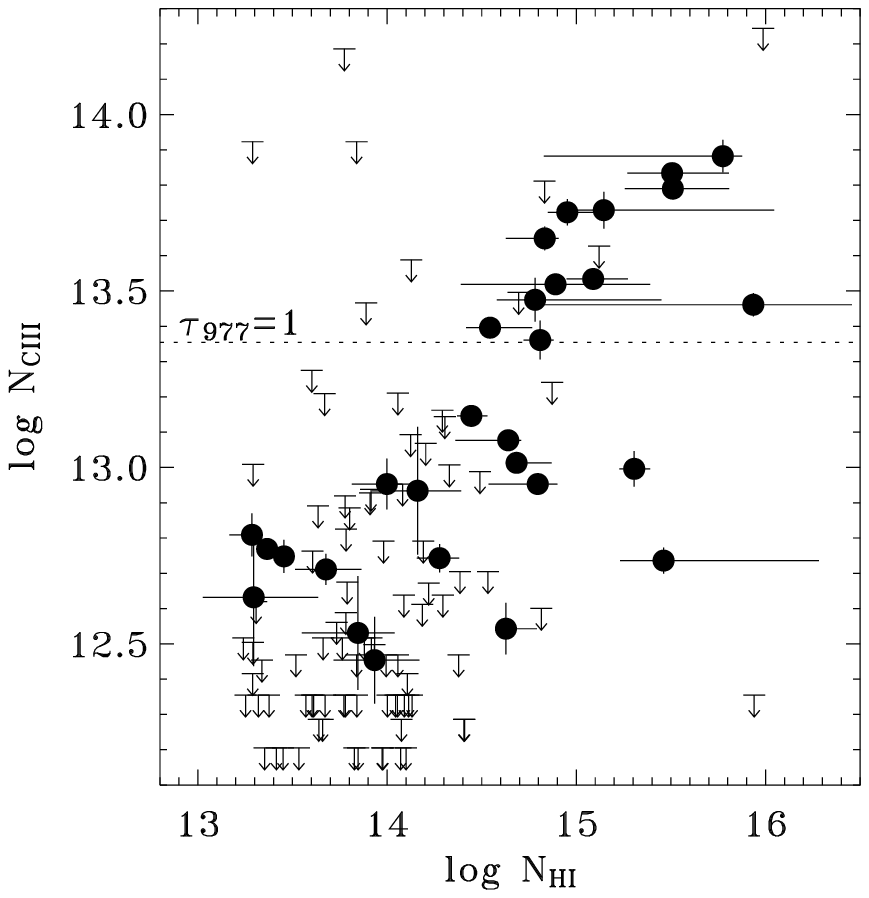}{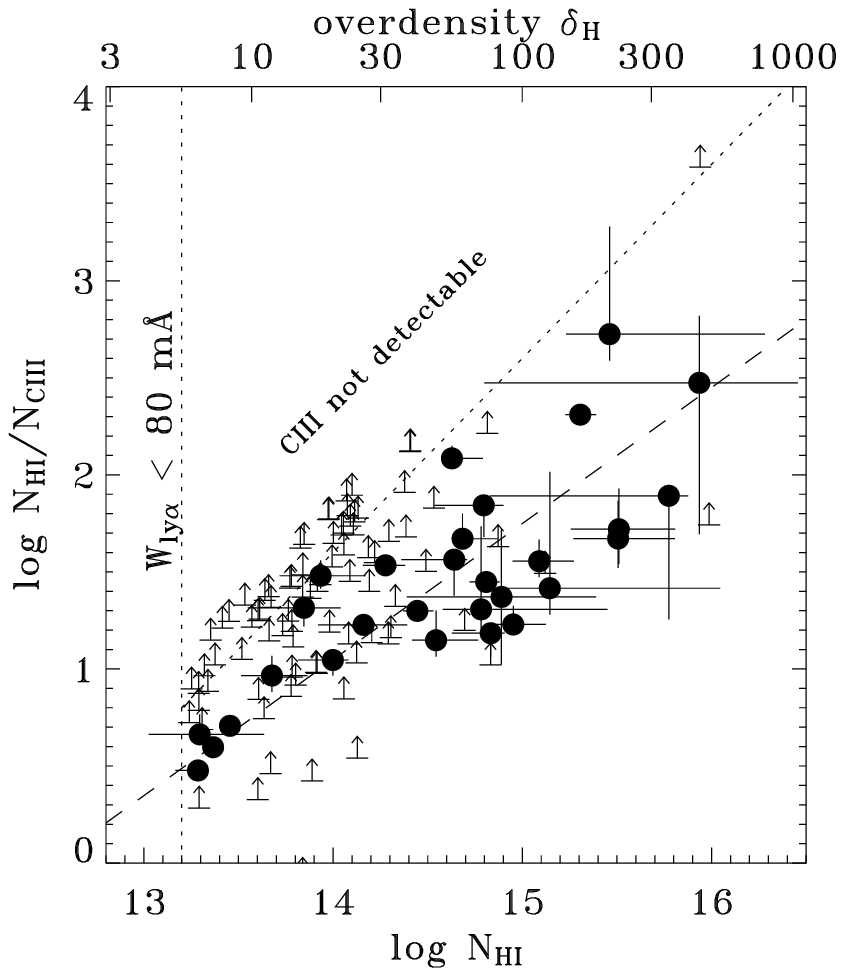} 
 \caption{Comparison of \NHI\ and \NCIII\ (left) and C\,III multiphase
 ratio vs.\ \NHI\ (right) for 30 IGM absorbers (circles) and 88 upper
 limits (arrows), similar to O\,VI analysis (Figure~1 of
 \citet{DanforthShull05}).  The H\,I and C\,III absorbers show a good
 correlation at low column density, implying that the ionization
 processes are coupled.  Higher column absorbers show considerably
 more scatter.  Overdensity $\delta_{\rm H}\equiv 20 N_{14}^{0.7}
 (10^{-0.4z})$ is shown on the top axis (see \citep[see][]{Dave99},
 and approximate detection limits for H\,I and C\,III are shown as
 dotted lines in the right panel.} \label{fig:nh1nc3}
\end{figure*}

The column density of \CIII\ shows a similar dynamic range (1.5 dex) to \NOVI\ (Figure~\ref{fig:nh1nc3}a).  Unlike \NOVI, \NCIII\ does show some degree of correlation with \NHI.  The \NHI/\NCIII\ multiphase ratio (Figure~\ref{fig:nh1nc3}b) is highly correlated in a similar manner as \NHI/\NOVI\ (Paper~I) for weak absorbers ($N_{\rm HI}\la10^{14.5}$ cm$^{-2}$), but the scatter becomes greater and the trend appears to level off for stronger absorbers.

It should be noted that a positive slope in multiphase ratio as a function of \NHI\ (Figure~\ref{fig:nh1nc3}b and Paper~I, Figure~1b) is {\em not} a metallicity effect (which would show a negative slope reflecting higher \HI\ fractions for higher-\NHI, metal-enriched gas).  We would also expect enriched material to be associated with galaxy outflows more than ``void'' IGM absorbers.  Stronger \lya\ absorbers are preferentially seen closer to galaxies than are weak absorbers \citep{Penton4}.

The lack of correlation in \OVI\ is a strong indicator that photoionization is not the principal ionizing mechanism of \OVI\ in the IGM.  If a species is the product of photoionization, its column densities should be roughly proportional to that of hydrogen in the optically thin limit.  We would expect a similar dynamic range and correlation in column density, seen as a horizontal trend in a multiphase plot.  \NCIII\ does show correlation with \NHI, but the dynamic ranges are still substantially different.

Photoionization may play a major role in \CIII\ production, but its signature proportionality to \HI\ and wide dynamic range in column density may be masked by the discrepancy between apparent and true column density (as discussed above for \lya\ lines).  \CIII\ $\lambda977$ is a strong transition and reaches $\tau_0=1$ at log\,$N_{\rm CIII}=13.35$ (for $b_{\rm CIII}=25$ \kms).  Forty percent of the \CIII\ detections are at or above this column density.  Conversely, the strongest \CIII\ absorbers show line-center apparent optical depths $\tau_{\rm a}\ga 2.0$.  The distribution of \NCIII\ may be truncated at higher columns by optical thickness, and the true column density distribution may be even shallower than shown in Figure~\ref{fig:c3plot}.

The \OVI\ doublet transitions are not as strong as \CIII\ $\lambda977$ and \OVI\ $\lambda1032$ reaches $\tau_0=1$ at log\,$N_{\rm OVI}=14.09$ (for $b_{\rm OVI}=25$ \kms), stronger than all but 12\% of the absorbers.  The highest apparent optical depth $\tau_{\rm a}$ at line center in our sample is $\tau_{\rm a}\approx1.5$ and most are at $\tau_{\rm a}<1$.  It seems safe to say that none of the \OVI\ absorbers are optically thick and that the observed upper limit on column density is not a saturation effect.  Therefore, we must be seeing some physical regulation mechanism on \NOVI.


\subsubsection{Single-Phase IGM Models}

\begin{figure*} 
 \plottwo{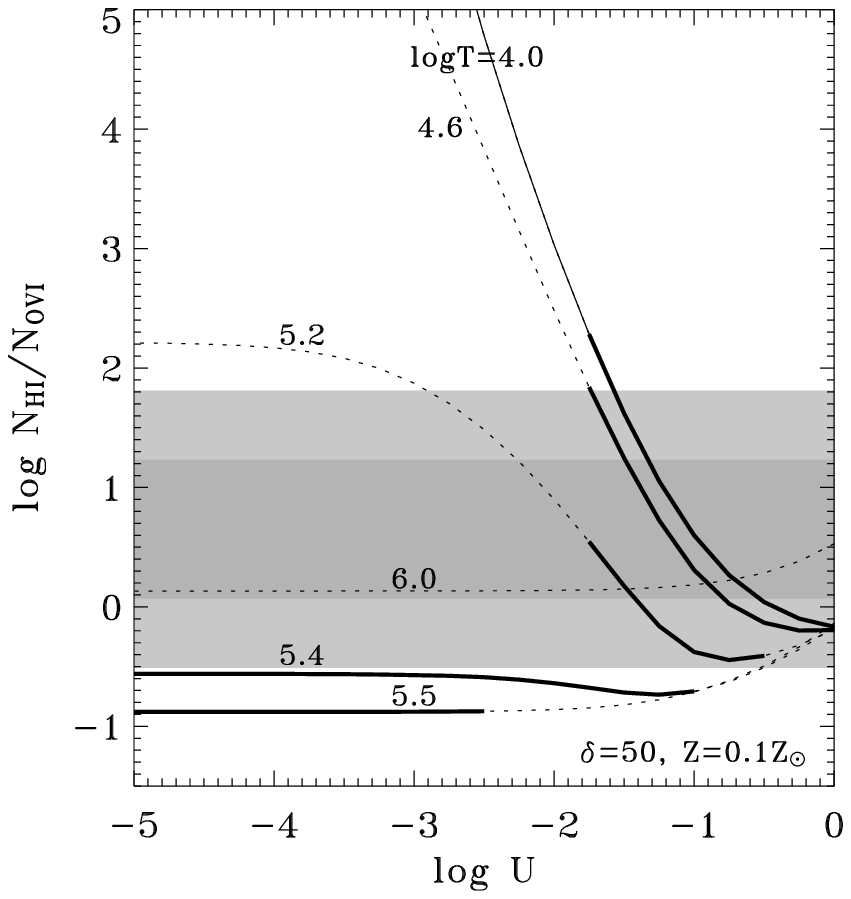}{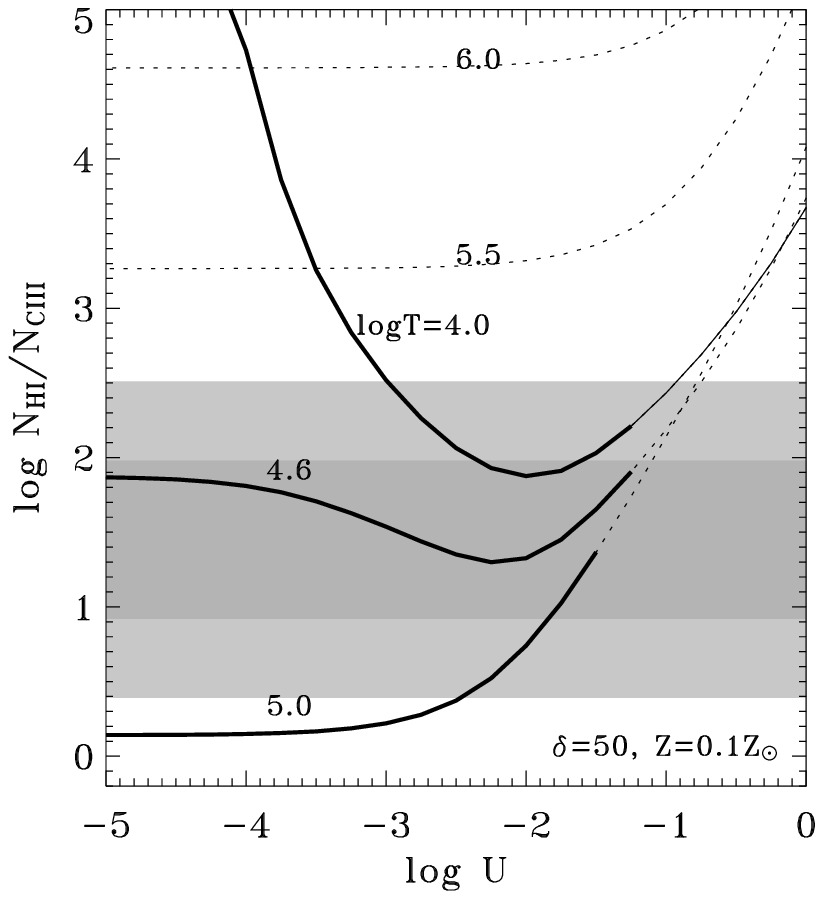} 
 \caption{The dependence of (\NHI/\NOVI)$_{\rm model}$ (left) and
 (\NHI/\NCIII)$_{\rm model}$ (right) on ionization parameter $U$ and
 temperature $T$ in {\sc CLOUDY} models of the IGM.  Models were
 calculated over a grid of ionization parameters, assuming a fixed
 temperature set by collisional heating.  In the left panel, we see
 the behavior of (\NHI/\NOVI)$_{\rm model}$ as a function of
 ionization parameter and temperature assuming $\delta=50$
 ($n_H=10^{-5}$ cm$^{-3}$) and $Z=0.1\,Z_\odot$.  The solid line
 represents the low-temperature case ($T=10^4$ K, equivalent to a
 pure photoionization model) and other curves represent
 photoionization plus steady-state temperature models of different
 temperatures.  Models with detectable columns of both H\,I and O\,VI
 (log\,$N_{\rm HI}>13.2$, log\,$N_{\rm OVI}>13.0$) are shown as
 thick, solid curves.  The shaded bar shows the $1\sigma$ (dark) and
 $2\sigma$ range of observed \NHI/\NOVI.  The right panel shows the
 equivalent plot for (\NHI/\NCIII)$_{\rm model}$ and observed values
 of \NHI/\NCIII.  From these, we can severely limit the allowed
 parameter space for a single-phase IGM absorber.}
 \label{fig:cloudymods} 
\end{figure*}

Even though we lack the range of ionic states to perform detailed comparisons with ionization models, we can still compare the range of observed values in a few lines with predicted behavior.  To this end, we constructed a grid of photoionization models with {\sc CLOUDY v96.01} \citep{CLOUDYref}.  An AGN ionizing continuum \citep{Korista97} illuminated a cloud 400 kpc thick at a distance of 10 Mpc from the absorber.  This scale is on the order of the scale predicted by our Hubble broadening arguments ($r\la300$ kpc) and $d{\cal N}/dz$ calculations presented below ($r\sim400$ kpc).  The number density $n_H$ and metallicity $Z$ of the cloud were varied between $10^{-6}$ cm$^{-2}$ and $10^{-4}$ cm$^{-3}$ ($\delta=5-500$ at $z\approx0$) and $Z=0.01-1\,Z_\odot$.  The luminosity of the illuminating AGN was varied to provide specified ionization parameters ($U=\phi/n_{\rm H} c$) log\,$U=-5$ to 0, equivalent to scaling the luminosity of the AGN, and the steady-state temperature was varied between $10^4$ K and $10^7$ K, the upper end of the WHIM temperature range.  The low end of the temperature scale is approximately equal to the photoexcitation temperature and thus approximates a pure-photoionization model similar to those of \citet{DonahueShull91}, while the low-ionization end of the grid approximates a purely collisional system in ionization equilibrium equivalent to \citet{SutherlandDopita93}.  In this way we simulated a single-phase system with both collisional and photoionization processes in effect.  Ion fractions for hydrogen, carbon, and oxygen were calculated as well as predicted column densities for \HI, \OVI, and \CIII.

Photoionization of \HI\ and \CIII\ depends on the shape of the ionizing spectrum in the 1-4 Ryd range, where the ambient radiation field is fairly well known.  However, \OVI\ is produced by photons of $E\ge114$ eV (8.4 Ryd) and typical AGN continua in this region are not as well determined.  In our models, we assume a power-law continuum in the EUV region; $I_\nu\propto\nu^{-\alpha}$.  The default CLOUDY AGN spectrum \citep{Korista97} uses $\alpha=1.4$, while recent studies find that the soft X-ray continuum follows a steeper power law with $\langle\alpha\rangle\sim1.8$.  We adopt $\alpha=1.8$ between 1 Ryd and 22 Ryd (300 eV) for our ionizing spectrum.  This provides less flux capable of photoionizing \OVI\ and increases the ionization parameters required to photoionize \OVI.

Figure~\ref{fig:cloudymods} shows how observed column density ratios (shaded ranges) compare with those from our model grid.  First we calculate the model multiphase ratios (\NHI/\NOVI)$_{\rm model}$ and (\NHI/\NCIII)$_{\rm model}$.  These are plotted as a function of $U$ for several different temperatures for the intermediate metallicity and density case: $\delta=50$ ($n_H=10^{-5}$ cm$^{-3}$) and $Z=0.1\,Z_\odot$.  Changes in metallicity move the curves up and down linearly.  Changes in number density have no effect on line ratios.  To further constrain the interpretation, we consider only models that produce column densities above the detection limits of this survey: log\,$N_{\rm HI}>13.2$, log\,$N_{\rm CIII}>12.5$, and log\,$N_{\rm OVI}>13.0$, shown in Figure~\ref{fig:cloudymods} as thick, solid lines.

\begin{figure*}
 \plotone{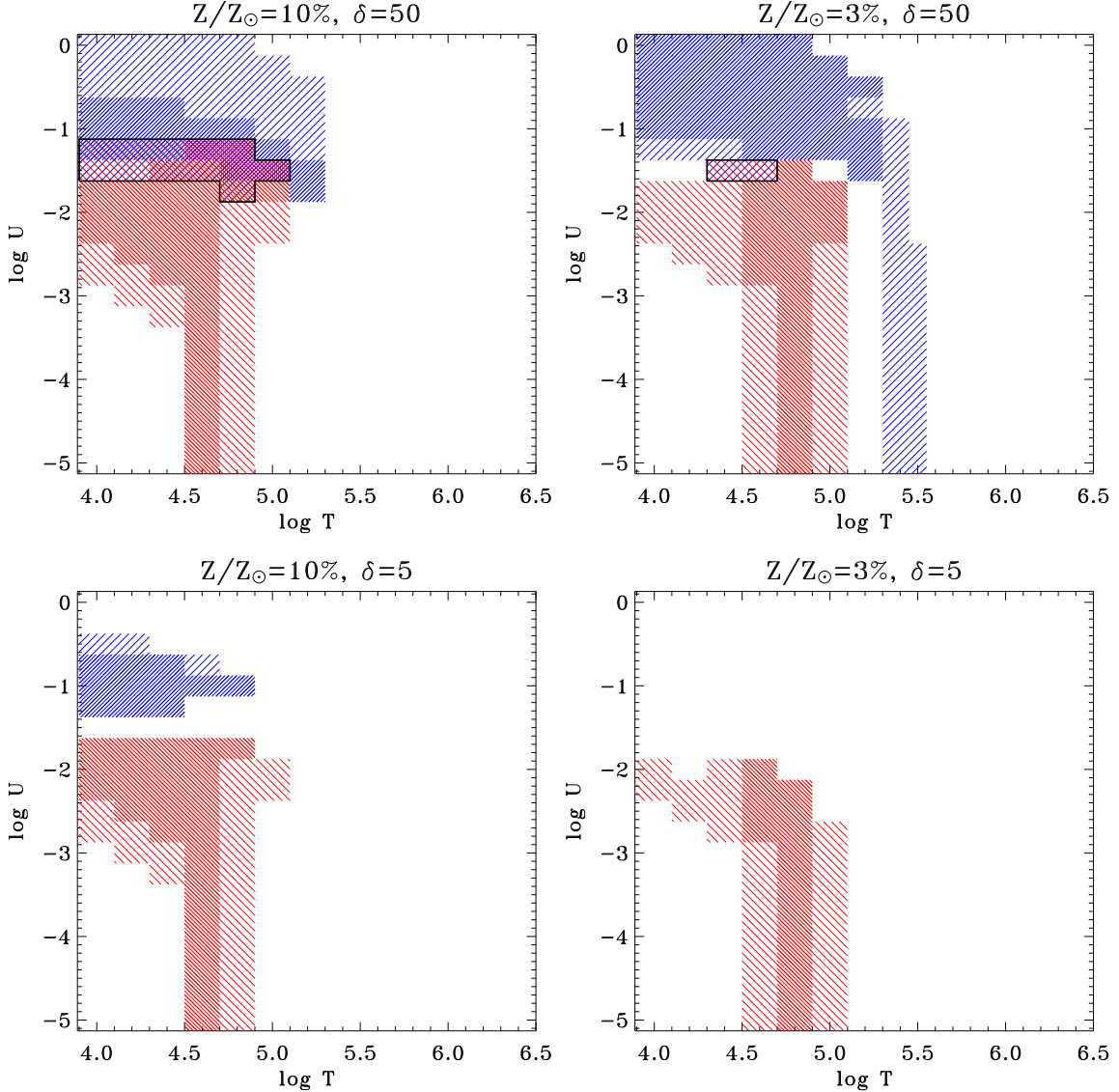} %
 \caption{Parameter space for which predicted column-density ratios
 fall within observed ranges.  Observationally allowed H\,I/O\,VI
 parameter space is shown in blue for the $\pm1\sigma$ (dense
 hatching) and $\pm2\sigma$ observed ranges.  H\,I/C\,III solutions
 are shown in red, and the overlap regions where both line ratios are
 within observed limits are noted with a black contour.  Model overdensity
 $\delta$ and metallicity are shown above each panel.  We see that
 the overlap region is small for a single-phase IGM model.}
 \label{fig:modelgrid} 
\end{figure*}

We see that several models generate multiphase ratios in the range of the observed quantities (shaded).  Either high-ionization (log\,$U\ga-2$) or very specific temperatures and low metallicity ($T=10^{5.4}-10^{5.5}$ K, $Z<0.1Z_\odot$) are required to match observed values of \NHI/\NOVI.  An ionization parameter of order $U\sim10^{-2}$ is feasible given a low-$z$ hydrogen ionizing flux $\Phi_{\rm ion}=3400$ photons cm$^{-2}$ s$^{-1}$ \citep{Shull99} and $n_{\rm H}=10^{-5}$ cm$^{-3}$, however the \NHI/\NCIII\ observations require low ionization and low temperatures (log\,$U\la-2$, $T\la10^5$~K).  No single-phase model can account for both the observed \OVI\ and \CIII\ absorbers.

To visualize this in another way, we show the area of $U$-$T$ parameter space for which (\NHI/\NOVI)$_{\rm model}$ and (\NHI/\NCIII)$_{\rm model}$ are within the observed ranges (Figure~\ref{fig:modelgrid}).  Models that produce \NHI/\NOVI\ within the observed range and which have column densities greater than the observed threshholds are shown as blue hatching.  Modeled values within $\pm1\sigma$ of the mean observed \NHI/\NOVI\ are shown in dense hatching, while those within $\pm2\sigma$ are shown in a less dense pattern.  The equivalent allowed parameter space for \NHI/\NCIII\ is shown in red.  There is very little overlap of the two allowed regions (denoted by black contours) in Figure~\ref{fig:modelgrid}.  A single-phase IGM with both \OVI\ and \CIII\ detections is unlikely if not actually impossible given our models.



The collisional ionization model is not without its own problems, however.  The time required to cool a diffuse plasma at WHIM temperatures can be approximated as 
\begin{equation} t_{\rm cool}\sim\frac{\frac{3}{2}k T}{n_{\rm H} \Lambda(T)} \sim (2.1~{\rm Gyr}) \frac{T_6}{n_{-4} \Lambda_{-22.5}} \; , \end{equation} 
where $T_6$ is the electron temperature in units of $10^6$~K, $n_{-4}$ is the electron density in units of $10^{-4}$ cm$^{-3}$, and $\Lambda_{-22.5}$ is the cooling rate coefficient in units of $10^{-22.5}$ erg cm$^3$ s$^{-1}$, typical of $Z=0.1-0.3 Z_\odot$ gas \citep{SutherlandDopita93}.  For $n=10^{-5}$~cm$^{-3}$ (an overdensity $\delta=50$ at $z=0$) and $Z=0.1~Z_{\odot}$, a $10^6$~K plasma will only cool in $\sim20$~Gyr.  Modifying the metallicity, density, or temperature will change $t_{\rm cool}$, but any reasonable set of IGM parameters will produce a minimum cooling time of a few Gyr.  Non-equilibrium calculations \citep{Rajan05,Furlanetto05} show qualitatively the same conclusion: hot gas at low densities stays hot for a very long time.  The frequency of shocking events (either cloud collisions or SN feedback shocks) required to maintain WHIM temperatures is fairly low.  Indeed, it becomes impressive that shock-heated IGM gas can cool enough to produce a significant \OVI\ fraction at all.  Shocks will increase $n_{\rm H}$ as well as $T$, but compression from an adiabatic shock can only account for a factor of 4 in density and only a small reduction in $t_{\rm cool}$.

The cooling is dominated (for $Z>10^{-2}Z_\odot$) by metal ion emission lines (primarily iron) at temperatures below $10^7$~K.  At lower temperatures, \CIII\ and \ion{C}{4} are some of the strongest coolants.  As temperatures reach $T<10^5$~K, \CIII\ becomes the dominant carbon ion and the gas cools rapidly.  Thus, \CIII\ is a transient ion in the cooling column of post-shock gas.  Because the shocked IGM takes a long time to cool to \CIII\ (and cools further to \ion{C}{2}), the presence of strong \CIII\ absorption suggests a source from photoionization, not from cooling shocks.  Certainly \OVI\ and \CIII\ are not simultaneously present in the same gas under CIE conditions.


\subsubsection{Multiphase IGM Models}

Single-phase models can potentially explain the absorbers for which we detect either \OVI\ or \CIII, but they cannot explain the twelve absorbers for which we see both \OVI\ {\em and} \CIII.  Is a multiphase, collisionally ionized system physically feasible? We postulate a shock generated by either a cloud collision (such as infall of material onto a filament) or a SNe-driven galactic outflow propagating through a metal-enriched intergalactic cloud.  We then investigate whether this model qualitatively reproduces the observed \OVI\ and \CIII\ detections.

Since $t_{\rm cool}$ is so long, we cannot assume that post-shock material will have a significant column of low ions.  Any \CIII\ and other low-ions must come from preshock gas.  If the crossing time is long enough, there can be unshocked gas available to provide the low ionization column seen in the observations.  We choose a shock velocity of $\sim200$ \kms\ and calculate the crossing time by estimating a typical IGM absorber scale.

We can estimate typical IGM absorber scale via low-$z$ detection statistics; 
\begin{equation} 
\frac{d{\cal N}}{dz}=~n_0(>L)~(\pi r_0^2)~\frac{c}{H_0} \; , 
\end{equation} 
where $n_0(>L)$ is the number density of galaxies brighter than a certain minimum luminosity and $\pi r_0^2$ is the typical absorber cross section.  We use a Schechter luminosity function 
\begin{equation} 
\phi(L)dL=\phi_*(L/L^*)^{-\alpha}~e^{-L/L^*}~(dL/L^*) \; , 
\end{equation} 
and integrate down to luminosity $L$.  In the special case of $\alpha=1$, typical of the faint-end slope, the integral becomes the first exponential integral $E_1$; 
\begin{eqnarray} 
n_0(>L) & = & \phi_*\int_L^\infty (L/L^*)^{-1}\,e^{-L/L^*}\,\frac{dL}{L^*}\nonumber\\
        & = & \phi_*\,E_1(L/L^*).  
\end{eqnarray} 
Recent results from the Sloan Digital Sky Survey \citep[SDSS][]{Blanton03} give $\phi_*=0.0149~h^3$~Mpc$^{-3}$ or $\phi_*=5.11\times10^{-3}~h_{70}^3$~Mpc$^{-3}$.  We use $d{\cal N}_{\rm OVI}/dz=17\pm3$ for $W_\lambda>30$ m\AA\ \citep{DanforthShull05} and find that $r_0=(1060\pm100)~h_{70}^{-1}$ kpc for $L^*$ galaxies.  However, it is likely that the IGM is enriched preferentially by smaller galaxies \citep{Stocke05}.  If we integrate down to $L=0.1L^*$, we find $r_0=(370\pm30)~h_{70}^{-1}$ kpc, a size scale in reasonable agreement with inferred distances of metal distributions from nearest-neighbor galaxies \citep{Stocke05}.  Adjusting $\phi_*$ downward by 1/3, to account for ellipticals that may not have outflows, we find that $r_0$ increases by $\sim20$\%.  This analysis assumes a uniform distribution of galaxies.  \citet{TumlinsonFang05} find similar scales with $r_0=750$ kpc and $r_0=300$ kpc for $L^*$ and $0.1L^*$ limits based on actual galaxy distributions from SDSS.  

A typical IGM absorber scale of $\sim400~h_{70}^{-1}$ kpc also agrees with \lya\ forest cloud sizes inferred from photoionization modeling \citep{Shull98,Schaye01,Tumlinson05} and with our rough upper limit on cloud scale based on Hubble broadening of \HI\ absorption lines.  Observations of quasar pairs give results not inconsistent with our derived value.  The quasar pair LBQS\,1343$+$264\,A/B shows characteristic $r_0\sim300~h_{70}^{-1}$ at $z\sim2$ \citep{Bechtold94,Dinshaw94}.  At lower redshift, \citet{Young01} find that a coherence length between \lya\ absorbers of 500--1000 kpc at $0.4<z<0.9$ in a rare triple QSO system.  

The crossing time for a 400 kpc cloud by a 200 \kms\ shock is 2 Gyr, which is of the same order as the WHIM cooling time.  Assuming that IGM absorbers are associated with dwarf galaxies and that shocking events are infrequent, it is perfectly feasible that unshocked, photoionized material would exist to provide the observed \CIII\ column density along AGN sight lines, while shocked material could provide the observed \OVI.  

We can now revisit the different samples of absorbers introduced in \S~4.3.  The 12 absorbers with both \OVI\ and \CIII\ detections can most plausibly be understood as multiphase absorbers.  Column densities and line widths show no correlation and are not easily explained with a single-phase photoionized-plus-collisional system.  Our sight lines pass through both quiescent photoionized and shocked regions, and we are unable to kinematically differentiate the two phases in the spectra.  The two metal ions occupy different temperature and ionization parameter regimes and are found in physically distinct parts of the absorber.  Likely the \HI\ absorption is associated with the cooler, photoionized phase, since $b_{\rm HI}\la40$ \kms\ in most cases.

There are eight absorbers with \HI\ and \OVI\ detections and good \CIII\ non-detections.  These systems may be large, diffuse, photoionized clouds with a high ionization parameter (log\,$U>-2$), sufficient to produce \OVI\ by photoionization, and high enough that carbon would be ionized to \ion{C}{4}.  The small fraction of neutral hydrogen at this high ionization parameter would show the narrow thermal profiles observed in our \HI\ sample.  Equivalently, these systems can be interpreted as two-phase systems with a shocked, WHIM phase (probed by \OVI) and an unshocked, photoionized phase observed in \HI.  The small range of \NOVI\ compared to \NHI\ and the lack of correlation between the two species in column density and $\beta$ suggests the latter, multiphase interpretation.  A broad, WHIM component in \HI\ may be masked by stronger narrow components or be below the detection threshhold of our data.

The four systems with \HI\ and \CIII\ detections and \OVI\ non-detections are likely photoionized.  CIE cooling is extremely fast at $T<10^5$~K, and we do not expect that $\sim10\%$ of the total metal absorbers would be observed during the relatively brief period during which \CIII\ is collisionally ionized.  The photoionization interpretation provides us with an upper limit to the ionization parameter ($U\la10^{-2}$) and we must therefore posit that, for this population of absorbers at least, $n_{\rm H}\gg10^{-5}$ cm$^{-3}$ or that the metagalactic ionizing radiation field is weaker than expected from models.

The 21 absorbers with neither \OVI\ nor \CIII\ are likely collisionally ionized to $T>10^6$~K or metal-poor systems ($Z<10^{-2}Z_\sun$).  The lack of broad \HI\ absorption makes the high-temperature interpretation implausible, and purely photoionized clouds with even modest enrichment should show measurable \CIII.  \citet{Stocke05} investigate our detection statistics more thoroughly and catalog nearest-neighbor galaxies for each absorber.  They find that the \lya\ detections with no metal lines tend to show larger nearest-neighbor distances than those with metal line detections.  \citet{Williger06} find stronger clustering between \lya\ systems and galaxies for higher \NHI\ systems over larger velocity scales.  These results suggest that the metal-enrichment explanation is the most likely.  


\subsection{Metallicity of the IGM}

In Paper~I, we found good agreement between our observed distribution of $d{\cal N}_{\rm OVI}/dz$ and the cosmic evolution models of \citet{Chen03} at $\sim10\%$ metallicity.  We also derived $Z_{\rm O}\approx0.09\,Z_\odot\,(f_{\rm OVI}/0.2)^{-1}$ based on the \NHI/\NOVI\ multiphase plot, where $(f_{\rm OVI}/0.2)$ is a normalized ionization fraction of \OVI\ in units of the CIE peak value of 20\%.  Our value of 9\% is consistent with the canonical 10\% value assumed in many sources \citep[e.g.,][]{Savage02,Tripp00}.

We determine $(C/H)_{\rm IGM}$ via the \HI/\CIII\ multiphase data using the same formalism as Paper~I,
\begin{equation}\label{eq:mainZ} 
 \biggl<\frac{N_{\rm C}}{N_{\rm H}}\biggr> = \biggl<\frac{N_{\rm CIII}}{N_{\rm HI}}\biggr> \times \biggl(\frac{f_{\rm HI}}{f_{\rm CIII}}\biggr) \; , 
\end{equation} 
where $f_{\rm HI}$ and $f_{\rm CIII}$ are the ionization fractions of those two species.  We fit the \CIII\ multiphase plot (Figure~\ref{fig:nh1nc3}b) as a power law 
\begin{equation} \label{eq:c3multiphasefit} 
 \biggl<\frac{N_{\rm HI}}{N_{\rm CIII}}\biggr> = C_{14}\,N_{14}^\alpha \; , 
\end{equation} 
where $N_{14}$ is the \HI\ column density in units of $10^{14}$ cm$^{-2}$ and $C_{14}$ is a scaling constant.  The best-fit parameters are similar, whether we use the low-\NHI\ half of the \CIII\ absorber sample or the entire range of \NHI:  for log\,$N_{\rm HI}<14.5$, $\alpha=0.73\pm0.08$, $C_{14}=1.06\pm0.04$; for the entire sample, $\alpha=0.70\pm0.03$, $C_{14}=1.05\pm0.02$.  We adopt the first set of values here, but note that it makes little difference.  The mean absorber redshift in our \CIII\ sample is $\langle z_{\rm abs}\rangle\approx0.1$.

We also make use of an empirical relationship from \citet{Dave99} relating the baryon overdensity, $\delta_{\rm H}$, to \HI\ column density 
\begin{eqnarray}
 \delta_{\rm H} &\equiv& \frac{n_H}{(1.90\times10^{-7}~{\rm cm^{-3}})\,(1+z)^3}\label{eq:deltadef} \\
 &\approx& 20\,N_{14}^{0.7}\,10^{-0.4\,z} \; .  \label{eq:deltadave} 
\end{eqnarray} 
Combining Eqs.~(\ref{eq:c3multiphasefit}) and (\ref{eq:deltadave}) and substituting the appropriate constants from the fit, we find 
\begin{equation} 
 \biggl<\frac{N_{\rm HI}}{N_{\rm CIII}}\biggr> = 26.6~\delta^{1.04} \; .  \label{eq:comb1} 
\end{equation}

The neutral hydrogen fraction $f_{\rm HI}$ in Eq.~\ref{eq:mainZ} can be derived from case-A photoionization equilibrium in a low density gas: 
\begin{eqnarray} 
 f_{\rm HI} &=& \frac{n_e \alpha_H^{(A)}}{\Gamma_H} \nonumber = 4.74\,n_H\,T_4^{-0.726}\,\Gamma_{-13}^{-1} \nonumber \\ 
  &=& 9.01\times10^{-7}\,\delta_{\rm H}\,(1+z)^3\,T_4^{-0.726}\,\Gamma_{-13}^{-1} \; , \label{eq:fh1} 
\end{eqnarray} 
where $T_4$ is the temperature in units of $10^4$~K and $\Gamma_{-13}$ is the \HI\ photoionization rate in units of $10^{-13} \rm~s^{-1}$.  The \CIII\ ionization fraction is harder to handle analytically.  We take $f_{\rm CIII}\approx0.8$ since this is roughly its maximum value under both CIE and the photoionization modeling of \citet{DonahueShull91}.  Substituting Eqs.~(\ref{eq:comb1}) and (\ref{eq:fh1}) into Eq~(\ref{eq:mainZ}), we get 
\begin{eqnarray} 
 \biggl<\frac{N_{\rm C}}{N_{\rm H}}\biggr> &=& (2.89\times10^{-5})\,\biggl(\frac{f_{\rm CIII}}{0.8}\biggr)^{-1}\,\delta_H^{-0.04}\,T_4^{-0.726}\,\Gamma_{-13}^{-1} \nonumber \\
 Z_{\rm C} & = & (0.12\,Z_\odot)\,\biggl(\frac{f_{\rm CIII}}{0.8}\biggr)^{-1}\; , 
\end{eqnarray} 
using $\rm(C/H)_\odot=2.45\times10^{-4}$ \citep{abundanceref}.  Our value $Z_{\rm C}=12\%$ is reassuringly close to the value $Z_{\rm O}=9\%$ from Paper~I, considering that both values are probably uncertain by at least a factor of two.  

The leading contender for IGM enrichment is outflows from starbursting dwarf galaxies \citep{Heckman01,Keeney05}.  These starburst winds would be dominated by the products of the most massive stars, which show an elevated abundance of oxygen in relation to carbon as a result of $\alpha$-process nucleosynthesis \citep{Garnett95,Sneden04}.  Within the large uncertainties, the observed $(C/O)_{\rm IGM}\approx(C/O)_\odot$ implies that the IGM is enriched by a more mature gas mixture from a broader range of stellar masses.  


\begin{figure}
  \epsscale{1.2}\plotone{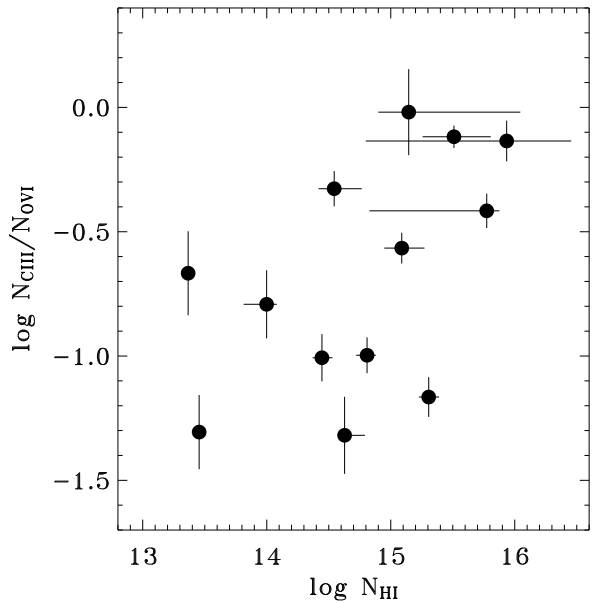} 
  \caption{The C\,III/O\,VI multiphase ratio shows a weak correlation
  with \NHI\ at the $2\sigma$ level ($R\sim0.6\pm0.2$).  The mean
  $N_{\rm CIII}/N_{\rm OVI}$ is consistent with
  $(C/O)=0.1^{+0.2}_{-0.1}~(C/O)_\odot$.}
  \label{fig:ionmultiphasevsh1}
\end{figure}

We can also directly compare measurements of O and C ions to determine (C/O)$_{\rm IGM}$.  Thirteen absorption systems show detections in both \OVI\ and \CIII\ with $\langle N_{\rm CIII}/N_{\rm OVI}\rangle=0.21^{+0.39}_{-0.14}$ ($1\sigma$ uncertainty, Figure~\ref{fig:ionmultiphasevsh1}).  We make some very crude assumptions for ionization fractions as discussed above, $f_{\rm CIII}=0.8$ and $f_{\rm OVI}=0.2$ based on peak CIE and/or photoionization values, and conclude that 
\begin{equation} 
 \biggl(\frac{C}{O}\biggr)_{\rm IGM}=0.1^{+0.2}_{-0.1}\,\biggl(\frac{C}{O}\biggr)_\odot\,\biggl(\frac{f_{\rm OVI}}{0.2}\biggr)\,\biggl(\frac{f_{\rm CIII}}{0.8}\biggr)^{-1} \; .  
\end{equation} 
There are large uncertainties, but this sub-solar abundance ratio is more consistent with what we expect from an enrichment of the IGM by the most massive stars.  However, it is in contradiction to the derived C/H and O/H metallicities from the multiphase plots discussed above and in Paper~I.

Given that \CIII\ and \OVI\ almost assuredly occupy different temperature phases in the IGM, as well as different spatial volumes within a given absorber (though they may share the same metallicity), these are not the best ions to use to investigate relative metallicities, so we view the $\rm(C/O)_{IGM}$ ratio above with caution.  Lithium-like \ion{C}{4} and \ion{N}{5} are more useful probes of relative abundances with \OVI.  These data are available in STIS spectra and will be the subject of a future investigation.


\section{Conclusions and Summary}

We present \FUSE\ observations of 31 AGN sight lines covering an integrated redshift path length of $\Delta z>2$.  We start with a known population of 171 \lya\ absorbers with $W_{\rm Ly\alpha}>80$~m\AA\ at $z\leq0.3$ and measure corresponding absorption in higher Lyman lines.  These allow us to determine $b_{\rm HI}$ and \NHI\ with a curve of growth, resulting in more accurate measurements than possible for \lya-only analysis.  

Higher Lyman lines are critical for accurate \NHI\ and $b_{\rm HI}$ measurements, particularly for stronger lines ($W_{\rm Ly\alpha}>180$~m\AA) where \lya\ absorption is saturated.  We find that $N_{\rm Ly\alpha}<N_{\rm COG}$ for most cases and that $b_{\rm Ly\alpha}$ consistently overpredicts $b_{\rm COG}$, often by a factor of two or more \citep[see also][]{Shull00}.

We measure corresponding metal-line absorption in \OVI\ (for $z\leq0.15$) and \CIII\ ($z<0.21$).  In all, \OVI\ is detected in 40 out of 129 possible absorbers, and \CIII\ is detected in 30 out of 148 absorbers.  \CIII\ detection statistics give $d{\cal N}_{\rm CIII}/dz=12^{+3}_{-2}$.  We calculate a typical size for IGM absorbers from detection statistics in \OVI\ and find $r_0 \sim 400$ kpc if absorbers are associated with $0.1L^*$ galaxies.  This size is similar to the results of \citet{TumlinsonFang05} and consistent with constraints placed on $r_0$ from Hubble broadening and observed $b$-values.

Our observations strongly suggest that \HI\ and \CIII\ probe photionized regions of the IGM, while \OVI\ is primarily a product of collisional (shock) ionization and therefore a valid probe of the WHIM.  Line width analysis sets an upper limit on absorber temperature.  All three species have similar distributions of $b$-values, with median values $\sim25$ \kms.  The \OVI\ is consistent with WHIM-phase, collisionally ionized WHIM, but $\langle b\rm(HI)\rangle=25\pm13$ \kms\ requires that the observed \HI\ absorption arise in a medium with $T<10^5$~K.  Single-phase CLOUDY models featuring both photoionization and collisional ionization limit which areas of parameter space are allowed for \OVI\ and \CIII\ absorbers.  Observed \NOVI\ values require log\,$U\ga-2$, while observed \NCIII\ requires log\,$U\la-2$, regardless of temperature.

We calculate the column density distribution of \HI\ absorbers and find that it follows a power law distribution ${\cal N}(N_{\rm HI})\propto N_{\rm HI}^{-\beta}$ with $\beta=1.68\pm0.11$, similar to the result found by \citet{Penton1,Penton4}.  We find that \CIII\ absorbers also follow a power-law distribution with slope $\beta_{\rm CIII}=1.68\pm0.04$, similar to $\beta_{\rm HI}$, but not as steep as $\beta_{\rm OVI} = 2.2 \pm 0.1$ found in Paper~I.  This similarity in $N$ distribution slope is circumstantial evidence that \HI\ and \CIII\ arise through similar mechanisms.

Our absorber sample includes 45 \HI\ absorbers with good statistics in both \OVI\ and \CIII\ lines.  We interpret 12 absorbers with \HI, \OVI, and \CIII\ as multi-phase systems with shock-heated WHIM (probed by \OVI) and a photoionized component seen in \CIII\ and \HI.  The four \HI+\CIII\ systems are interpreted as unshocked, photoionized gas at $10^4$~K.  The eight systems with \HI\ and \OVI\ but no \CIII\ are probably multi-phase absorbers with photoionized \HI\ and WHIM \OVI.  The 21 absorbers with neither \OVI\ nor \CIII\ are most likely low-metallicity, unshocked systems.  \citet{Stocke05} find that these systems have larger nearest-neighbor distances than the population of metal absorbers.

Finally, the metallicity of IGM absorbers appears close to the canonical 10\% solar value.  The \NHI/\NCIII\ multiphase relationship implies an IGM carbon metallicity $Z_C=0.12\,Z_\sun\,(f_{\rm CIII}/0.8)^{-1}$, remarkably consistent with the result $Z_O=0.09\,Z_\sun\,(f_{\rm OVI}/0.2)^{-1}$ from Paper~I.  This implies a near-solar abundance ratio of C/O in the IGM, higher than what we expect from stellar nucleosynthesis models which say that the IGM is preferentially enriched by O-rich massive stars.

Future work with other highly ionized species (e.g. \ion{C}{4}, \ion{Si}{3}, and \ion{Si}{4}) will help constrain ionization and temperature and cast more light on the metallicity trends in the low-redshift universe.  Deeper FUV observations will allow us to seach for lower column density absorbers, and additional sight lines from future observations will contribute redshift pathlength to the statistics.

\vspace{0.5cm}

It is our pleasure to acknowledge useful discussions with Nahum Arav, Jack Gabel, and Van Dixon.  Steve Penton reduced the STIS/E140M data for six sight lines.  We made extensive use of {\sc CLOUDY} v.96.01 and are grateful to Gary Ferland and Peter van Hoof for technical assistance with the code.  We would also like to thank Gerry Williger for a thorough job refereeing our manuscript.  This work contains data obtained for the Guaranteed Time Team by the NASA-CNES-CSA \FUSE\ mission operated by the Johns Hopkins University, as well as data from the {\it Hubble Space Telescope}.  J.L.R. has received financial support from NSF grant AST-0302049.  Financial support to the University of Colorado has been provided by NASA/\FUSE\ contract NAS5-32985 and grant NAG5-13004, by our HST \lya\ survey (GO Program 6593), and by theoretical grants from NASA/LTSA (NAG5-7262)and NSF (AST02-06042).

\LongTables
\begin{deluxetable}{lclrrrl}
\tabletypesize{\footnotesize}
\tablecolumns{7} 
\tablewidth{0pt} 
\tablecaption{Ly$\alpha$ Absorbers}
\tablehead{\colhead{Sight Line}                &
	   \colhead{Abs}                       &
           \colhead{$z\rm_{abs}$}              &
           \colhead{$v\rm_{lsr}$}              &
	   \colhead{$b$}                       & 
           \colhead{$W\rm_{rest}$}             &
	   \colhead{Source and Notes}          \\
	   \colhead{}                          &
	   \colhead{\#}                        &
           \colhead{}                          &
	   \colhead{(km/s)}                    &
           \colhead{(km/s)}                    &
           \colhead{(m\AA)}                    &
	   \colhead{}                          }
\startdata
 Mrk335      & 1&  0.00655&$  1965\pm ~6$&$    75\pm ~7$&$  229\pm ~30$& \citet{Penton1}  \\
 Mrk335      & 2&  0.00765&$  2295\pm 12$&$    73\pm 17$&$   81\pm ~26$& \citet{Penton1}  \\
 Mrk335      & 3&  0.02090&$  6269\pm ~6$&$    75\pm ~6$&$  130\pm ~14$& \citet{Penton1}  \\
 IZw1        & 1&  0.00539&$  1617\pm ~5$&   \nodata    &$  120\pm ~37$& \citet{Shull96}  \\
 IZw1        & 2&  0.00954&$  2861\pm ~9$&   \nodata    &$   84\pm ~40$& \citet{Shull96}  \\
 IZw1        & 3&  0.01710&$  5130\pm 12$&   \nodata    &$   84\pm ~47$& \citet{Shull96}  \\
 TonS180     & 1&  0.01834&$  5502\pm 11$&$    51\pm ~5$&$  268\pm ~54$& \citet{Penton4}  \\
 TonS180     & 2&  0.02339&$  7017\pm 11$&$    56\pm ~4$&$  222\pm ~29$& \citet{Penton4}  \\
\enddata		   	  
\end{deluxetable}

\clearpage
\LongTables
\begin{deluxetable}{lclllll}
\tabletypesize{\footnotesize}
\tablecolumns{7} 
\tablewidth{0pt} 
\tablecaption{H\,I Absorber Measurements}
\tablehead{\colhead{Target}            &
   \colhead{Abs\#}                     &
   \colhead{$z\rm_{abs}$}              &
   \colhead{$cz_{\rm abs}$}            &
   \colhead{$b_{\rm HI}$}              &
   \colhead{log\,\NHI}                 &
   \colhead{notes\tablenotemark{a}} }
\startdata
Mrk335      & 1&0.0066&  1965&$>43               $&$ 13.611^{+0.062}_{-0.020}$&this work\\
Mrk335      & 2&0.0077&  2295&$17^{+\infty}_{-13}$&$ 13.295^{+0.341}_{-0.269}$&this work\\
Mrk335      & 3&0.0209&  6269&$25^{+\infty}_{-14}$&$ 13.518^{+0.200}_{-0.149}$&this work\\
IZw1        & 1&0.0054&  1617&$20\pm10           $&$ 13.500^{+0.292}_{-0.062}$&$W_{\rm Ly\alpha}$ \tablenotemark{b}\\
IZw1        & 2&0.0095&  2861&$20\pm10           $&$ 13.289^{+0.137}_{-0.039}$&$W_{\rm Ly\alpha}$ \tablenotemark{b}\\
IZw1        & 3&0.0171&  5130&$20\pm10           $&$ 13.289^{+0.137}_{-0.039}$&$W_{\rm Ly\alpha}$ \tablenotemark{b}\\
TonS180     & 1&0.0183&  5502&$30^{+\infty}_{-12}$&$ 13.992^{+0.178}_{-0.143}$&this work\\
TonS180     & 2&0.0234&  7017&$16^{+ 6}_{- 5}    $&$ 14.385^{+0.522}_{-0.234}$&this work\\
\enddata
\tablenotetext{a}{Unless otherwise noted, quoted $b_{\rm HI}$ and log\,$N_{\rm HI}$ values are based on concordance plots using multiple Lyman lines.  ``\lya\ fit'' denotes cases based on Voigt fits to \lya\ only.}
\tablenotetext{b}{Column density estimate based on $W_{\rm Ly\alpha}$ and assumed $b=20\pm10$ \kms.}
\tablenotetext{c}{Deviation of $>2\sigma$ in $b_{\rm HI}$ compared with published, this work-qualified measurement.}
\tablenotetext{d}{Deviation of $>2\sigma$ in $N_{\rm HI}$ compared with published, this work-qualified measurement.}
\end{deluxetable}
\clearpage
\LongTables
\begin{deluxetable}{lclrcccccl}
\tabletypesize{\footnotesize}
\tablecolumns{10} 
\tablewidth{0pt} 
\tablecaption{O\,VI Absorber Detections and Upper Limits}
\tablehead{\colhead{Sight Line}        &
   \colhead{Abs}                       &
   \colhead{$z\rm_{abs}$}              &
   \colhead{$cz_{\rm abs}$}            &
   \colhead{$b_{1032}$}                &
   \colhead{$W_{1032}$}                &
   \colhead{$b_{1032}$}                &
   \colhead{$W_{1038}$}                &
   \colhead{log\,N(OVI)}               &
   \colhead{notes} \\
   \colhead{}&
   \colhead{\#}&
   \colhead{}&
   \colhead{(\kms)}&
   \colhead{(\kms)}&
   \colhead{(m\AA)}&
   \colhead{(\kms)}&
   \colhead{(m\AA)}&
   \colhead{}&
   \colhead{}
}
\startdata
Mrk335      & 1&0.0066&  1965&\nodata  &\nodata    &\nodata  &$< 10     $&$<13.21        $&\\
Mrk335      & 2&0.0077&  2295&\nodata  &$<  9     $&\nodata  &$< 10     $&$<12.87        $&\\
Mrk335      & 3&0.0209&  6269&\nodata  &$< 31     $&\nodata  &$< 17     $&$<13.42        $&\\
IZw1        & 2&0.0095&  2861&\nodata  &$< 42     $&\nodata  &$< 39     $&$<13.57        $&\\
TonS180     & 1&0.0183&  5502&\nodata  &$< 20     $&\nodata  &$< 20     $&$<13.22        $&\\
TonS180     & 2&0.0234&  7017& 30      &$ 43\pm 15$&\nodata  &$< 20     $&$ 13.48\pm 0.05$&\\
TonS180     & 3&0.0430& 12912&\nodata  &$< 17     $&\nodata  &$< 18     $&$<13.15        $&\\
TonS180     & 4&0.0436& 13068& 40      &$ 31\pm 18$&\nodata  &\nodata    &$ 13.43\pm 0.02$&\\
\enddata
\end{deluxetable}
\clearpage
\LongTables
\begin{deluxetable}{lclrcccl}
\tabletypesize{\footnotesize}
\tablecolumns{8} 
\tablewidth{0pt} 
\tablecaption{C\,III Absorber Detections and Upper Limits}
\tablehead{\colhead{Sight Line}            &
   \colhead{Abs}                       &
   \colhead{$z\rm_{abs}$}              &
   \colhead{$cz_{\rm abs}$}            &
   \colhead{$b_{977}$}                 &
   \colhead{$W_{977}$}                 &
   \colhead{log\,N(CIII)}              &
   \colhead{notes} \\
   \colhead{}&
   \colhead{\#}&
   \colhead{}&
   \colhead{(\kms)}&
   \colhead{(\kms)}&
   \colhead{(m\AA)}&
   \colhead{}&
   \colhead{}
 }
\startdata
Mrk335      & 2&0.0077&  2295& 13      &$ 25\pm 13$&$ 12.63\pm 0.19$&\\
Mrk335      & 3&0.0209&  6269&\nodata  &$< 18     $&$<12.47        $&\\
IZw1        & 3&0.0171&  5130&\nodata  &$<701     $&$<13.92        $&\\
TonS180     & 2&0.0234&  7017&\nodata  &$< 30     $&$<12.70        $&\\
TonS180     & 3&0.0430& 12912&\nodata  &$< 21     $&$<12.52        $&\\
TonS180     & 4&0.0436& 13068&\nodata  &$< 48     $&$<12.95        $&\\
TonS180     & 5&0.0450& 13515&\nodata  &$< 21     $&$<12.52        $&\\
TonS180     & 6&0.0456& 13681& 13      &$ 48\pm 12$&$ 12.95\pm 0.07$&\\
\enddata
\tablenotetext{a}{Absorber shows clear multiple structure.  Quoted values are sums of individually fit components.}
\tablenotetext{b}{Strong absorber associated with Lyman limit system.}
\end{deluxetable}


\begin{thebibliography}

\bibitem[Allende Prieto et al.(2001a)]{abundanceref} 
Allende Prieto, C., Lambert, D.  L., \& Asplund, M. 2001, \apjl, 556, L63
\bibitem[Allende Prieto et al.(2001b)]{Allende01} 
Allende Prieto, C., Lambert, D. L., \& Asplund, M. 2001, \apjl, 573, L137 
\bibitem[Bechtold et al.(1994)]{Bechtold94} 
Bechtold, J. B., Crotts, A. P. S., Duncan, R. C., \& Fang, Y. 1994, \apj, 437, L83
\bibitem[Bechtold et al.(2002)]{Bechtold02} 
Bechtold, J. B., et al. 2002, \apjs, 140, 143
\bibitem[Blanton et al.(2003)]{Blanton03} 
Blanton, M. R., et al. 2003, \apj, 592, 819
\bibitem[Birnboim \& Dekel(2002)]{BirnboimDekel02} 
Birnboim, Y. \& Dekel, A. 2002, \mnras, 345, 349
\bibitem[Cen \& Ostriker(1999a)]{CenOstriker99a} 
Cen, R., \& Ostriker, J. P. 1999a, \apj, 519, L109 
\bibitem[Cen \& Ostriker(1999b)]{CenOstriker99b} 
Cen, R., \& Ostriker, J. P. 1999b, \apj, 514, 1 
\bibitem[Cen et al.(2001)]{CenOstriker01} 
Cen, R., Tripp, T. M., Ostriker, J. P., \& Jenkins, E. 2001, \apj, 559, L5
\bibitem[Chen et al.(2003)]{Chen03} 
Chen, X., Weinberg, D. H., Katz, N., \& Dav\'e, R. 2003, \apj, 594, 42
\bibitem[Collins, Shull, \& Giroux(2004)]{Collins04} 
Collins, J. A., Shull, J. M., \& Giroux, M. L. 2004, \apj, 605, 216
\bibitem[Crenshaw et al.(1999)]{Crenshaw99} 
Crenshaw, D. M., Kraemer, S. B., Boggess, A., Maran, S. P., Mushotzky, R. F., \& Wu, C. 1999, \apj, 516, 750
\bibitem[Danforth \& Shull(2005)]{DanforthShull05} 
Danforth, C. W. \& Shull, J. M. 2005, \apj, 624, 555 (Paper~I)
\bibitem[Dav\'e et al.(1999)]{Dave99} 
Dav\'e, R., et al. 1999, \apj, 511, 521 
\bibitem[Dav\'e et al.(2001)]{Dave01} 
Dav\'e, R., et al. 2001, \apj, 552, 473 
\bibitem[Dav\'e \& Tripp(2001)]{DaveTripp01} 
Dav\'e, R., \& Tripp, T. M. 2001, \apj, 553, 528 
\bibitem[Dinshaw et al.(1994)]{Dinshaw94} 
Dinshaw, N., Weymann, R. J., Impey, C. D., Foltz, C. B., Morris, S. L., \& Ake, T. 1997, \apj, 437, L87
\bibitem[Donahue \& Shull(1991)]{DonahueShull91} 
Donahue, M. \& Shull, J. M. 1991, \apj, 383, 511
\bibitem[Ferland et al.(1998)]{CLOUDYref} 
Ferland, G. J., et al. 1998, \pasp, 110, 761
\bibitem[Furlanetto, Phillips, \& Kamionkowski(2005)]{Furlanetto05} 
Furlanetto, S. R., Phillips, L. A., \& Kamionkowski, M. 2005, \mnras, 359, 295
\bibitem[Garnett et al.(1995)]{Garnett95} 
Garnett, D. R., et al. 1995, \apj, 443, 64
\bibitem[Gehrels(1986)]{Gehrels86} 
Gehrels, N. 1986, \apj, 303, 336 
\bibitem[Gillmon et al.(2005)]{Gillmon05} 
Gillmon, K., Shull, J. M., Tumlinson, J., \& Danforth, C. W. 2005, \apj, submitted (astro-ph/0507581) 
\bibitem[Hartigan, Raymond, \& Hartmann(1987)]{HRH87} 
Hartigan, P., Raymond, J., \& Hartmann, L.\ 1987, \apj, 316, 323 
\bibitem[Hebrard et al.(2006)]{Hebrard05} 
Hebrard, G., et al. 2006, \apj, accepted (astro-ph/0508611)
\bibitem[Heckman et al.(2001)]{Heckman01} 
Heckman, T. M., et al. 2001, \apj, 554, 1021
\bibitem[Heckman et al.(2002)]{Heckman02} 
Heckman, T. M., Norman, C. A., Strickland, D. K., \& Sembach, K. R. 2002, \apj, 577, 691
\bibitem[Hu et al.(1995)]{Hu95} 
Hu, E. M., Kim, T.-S., Cowie, L. L., Songaila, A., \& Rauch, M. 1995, \aj, 110, 1526
\bibitem[Jenkins(1986)]{Jenkins86} 
Jenkins, E. B. 1986, \apj, 304, 739
\bibitem[Jenkins et al.(2003)]{Jenkins03} 
Jenkins, E. B., et al. 2003, \aj, 125, 2824
\bibitem[Keeney et al.(2006)]{Keeney05} 
Keeney, B., Danforth, C. W., Stocke, J. T., Penton, S. V., \& Shull, J. M. 2006, in prep
\bibitem[Kim et al.(1997)]{Kim97} 
Kim, T.-S., Hu, E. M., Cowie, L. L., \& Songaila, A. 1997, \aj, 114, 1
\bibitem[Kriss(2002)]{Kriss02} 
Kriss, G. A. 2002, in ``Mass Outflow in Active Galactic Nuclei: New Perspectives'', ASP Conf. Ser. 255, 69
\bibitem[Korista et al.(1997)]{Korista97} 
Korista, K., et al. 1997, \apjs, 108, 401
\bibitem[Lopez et al.(1999)]{Lopez99} 
Lopez, S., Reimers, D., Rauch, M., Sargent, W. L., \& Smette, A. 1999, \apj, 513, 598
\bibitem[Lu et al.(1996)]{Lu96} 
Lu, L., Sargent, W. L. W., Womble, D. S., \& Takada-Hidai, M. 1996, \apj, 472, 509
\bibitem[Mathews \& Ferland(1987)]{MathewsFerland87} 
Mathews, W. D., \& Ferland, G. 1987, \apj, 323, 456
\bibitem[Moos et~al.(2000)]{Moos00}
Moos, H. W., et~al. 2000, \apj, 538, L1
\bibitem[Parnell \& Carswell(1988)]{ParnellCarswell88} 
Parnell, H. C. \& Carswell, R.~F. 1988, \mnras, 230, 491
\bibitem[Penton, Stocke, \& Shull(2000)]{Penton1} 
Penton, S.~V., Stocke, J.~T., \& Shull, J.~M. 2000, \apjs, 130, 121
\bibitem[Penton, Shull, \& Stocke(2000)]{Penton2} 
Penton, S.~V., Shull, J.~M., \& Stocke, J.~T. 2000, \apj, 544, 150 
\bibitem[Penton, Stocke, \& Shull(2003)]{Penton3} 
Penton, S.~V., Stocke, J.~T., \& Shull, J.~M. 2003, \apj, 565, 720
\bibitem[Penton, Stocke, \& Shull(2004)]{Penton4} 
Penton, S.~V., Stocke, J.~T., \& Shull, J.~M. 2004, \apjs, 152, 29
\bibitem[Prochaska et al.(2004)]{Prochaska04} 
Prochaska, J. X., Chen, H-W, Howk, J. C., Weiner, B. J., \& Mulchaey, J. 2004, \apj, 617, 718
\bibitem[Rajan \& Shull(2005)]{Rajan05} 
Rajan, N. \& Shull, J. M. 2005, in prep.
\bibitem[Richter et al.(2004)]{Richter04} 
Richter, P., Savage, B. D., Tripp, T. M., \& Sembach, K. R. 2004, \apjs, 153, 165
\bibitem[Richter et al.(2005)]{Richter05} 
Richter, P., Savage, B. D., Tripp, T. M., \& Sembach, K. R. 2005, Proceedings of Science, ``Baryons in Dark Matter Halos'', Novigrad, Croatia (astro-ph/0412133)
\bibitem[Sahnow et~al.(2000)]{Sahnow00} 
Sahnow, D. J., et~al. 2000, \apj, 538, L7
\bibitem[Savage \& Sembach(1991)]{SavageSembach91} 
Savage, B.~D., \& Sembach, K.~R. 1991, \apj, 379, 245 
\bibitem[Savage et al.(2002)]{Savage02} 
Savage, B.~D., Sembach, K.~R., Tripp, T.~M., \& Richter, P. 2002, \apj, 564, 631
\bibitem[Schaye(2001)]{Schaye01} 
Schaye, J. 2001, \apj, 559, 507
\bibitem[Sembach et al.(2001)]{Sembach01} 
Sembach, K. R., Howk, J. C., Savage, B. D., Shull, J. M., \& Oegerle, W. R. 2001, \apj, 561, 573
\bibitem[Sembach et al.(2004)]{Sembach04} 
Sembach, K. R., Tripp, T. M., Savage, B. D., \& Richter, P. 2004, \apjs, 153, 165
\bibitem[Shull et al.(1998)]{Shull98} 
Shull, J. M., et al. 1998, \aj, 116 2094
\bibitem[Shull et al.(1999)]{Shull99} 
Shull, J. M., Roberts, D., Giroux, M. L., Penton, S. V., \& Fardal, M. A. 1999, \apj, 118, 1450
\bibitem[Shull et al.(2000)]{Shull00} 
Shull, J. M., et al. 2000, \apj, 538, L13 
\bibitem[Shull, Stocke, \& Penton(1996)]{Shull96} 
Shull, J. M., Stocke, J. T., \& Penton, S. V. 1996, \aj, 111, 72
\bibitem[Shull, Tumlinson, \& Giroux(2003)]{Shull03} 
Shull, J. M., Tumlinson, J., \& Giroux, M. 2003, ApJ, 594, L107 
\bibitem[Sneden(2004)]{Sneden04} 
Sneden, C. 2004, Mem. S. A. It., 75, 267
\bibitem[Stocke, Shull, \& Penton(2005)]{Stocke04} 
Stocke, J. T., Shull, J. M., \& Penton, S. V. 2005, in ``From Planets to Cosmology: Proceedings of STScI May 2004 Symposium'', in press (astro-ph/0407352)
\bibitem[Stocke et al.(2005)]{Stocke05} 
Stocke, J. T., Penton, S. V., Danforth, C. W., Shull, J. M., Tumlinson, J., \& McLin, K. 2005, \apj, submitted
\bibitem[Sutherland \& Dopita(1993)]{SutherlandDopita93} 
Sutherland, R. S., \& Dopita, M. A. 1993, \apjs, 88, 253
\bibitem[Tripp, Lu, \& Savage(1998)]{Tripp98} 
Tripp, T. M., Lu, L., \& Savage, B. D. 1998, \apj, 508, 200
\bibitem[Tripp, Savage, \& Jenkins(2000)]{Tripp00} 
Tripp, T. M., Savage, B. D., \& Jenkins, E. B. 2000, \apj, 534, L1
\bibitem[Tumlinson et al.(2005)]{Tumlinson05} 
Tumlinson, J., Shull, J. M., Giroux, M. L., \& Stocke, J. T. 2005, \apj, 620, 95
\bibitem[Tumlinson \& Fang(2005)]{TumlinsonFang05} 
Tumlinson, J. \& Fang, T. 2005, \apj, 623, L97
\bibitem[Weinberg, Katz, \& Hernquist(1998)]{Weinberg98} 
Weinberg, D. H., Katz, N., \& Hernquist, L. 1998, in ASP Conf. Ser. 148, Origins, ed. C. E. Woodward, J. M. Shull, \& H. A. Thronson (San Francisco,: ASP), 21
\bibitem[Weymann et al.(1998)]{Weymann98} 
Weymann, R., et al. 1998, \apj, 506, 1
\bibitem[Williger et al.(2005)]{Williger05} 
Williger, G. M., et al. 2005, \apj, 625, 210
\bibitem[Williger et al.(2006)]{Williger06} 
Williger, G. M., et al. \apj, in press, 10 Jan. 2006 (astro-ph/0505586)
\bibitem[Young, Impey, \& Foltz(2001)]{Young01} 
Young, P. A., Impey, C. D., \& Foltz, C. B. 2001, \apj, 549, 76
\end{thebibliography}
\end{document}